\documentclass{aa}

\usepackage{graphicx}
\usepackage{txfonts}
\usepackage{color}

\begin{document}

\title{Chemical enrichment and accretion of nitrogen-loud quasars\thanks{Based on observations collected at the European Organisation for Astronomical Research in the Southern Hemisphere under ESO programme 088.B-0191(A), and at the Subaru Telescope, which is operated by the National Astronomical Observatory of Japan.}}

\subtitle{}

\author{K. Matsuoka\inst{1,2}
        \and
        T. Nagao\inst{3}
        \and
        R. Maiolino\inst{4,5}
        \and
        A. Marconi\inst{1}
        \and
        D. Park\inst{6}
        \and
        Y. Taniguchi\inst{7}
}

\institute{Dipartimento di Fisica e Astronomia, Universit{\'a} degli Studi di Firenze,
           Via G. Sansone 1, I-50019 Sesto Fiorentino, Italy
           \and
           INAF -- Osservatorio Astrofisico di Arcetri,
           Largo Enrico Fermi 5, I-50125 Firenze, Italy\\
           \email{matsuoka@arcetri.astro.it}
           \and
           Research Center for Space and Cosmic Evolution,
           Ehime University, 2-5 Bunkyo-cho, Matsuyama 790-8577, Japan
           \and
           Cavendish Laboratory, University of Cambridge,
           19 J. J. Thomson Avenue, Cambridge CB3 0HE, UK
           \and
           Kavli Institute for Cosmology, University of Cambridge,
           Madingley Road, Cambridge CB3 0HA, UK
           \and
           Korea Astronomy and Space Science Institute,
           776 Daedeok-daero, Yuseong-gu, Daejeon 34055, Republic of Korea
           \and
           The Open University of Japan,
           2-11 Wakaba, Mihama-ku, Chiba 261-8586, Japan
}

\date{}

\abstract{We present rest-frame optical spectra of 12 ``nitrogen-loud'' quasars at $z \sim 2.2$, whose rest-frame ultraviolet (UV) spectra show strong nitrogen broad emission lines. To investigate their narrow-line region (NLR) metallicities, we measure the equivalent width (EW) of the [\ion{O}{iii}]$\lambda$5007 emission line: if the NLR metallicity is remarkably high as suggested by strong UV nitrogen lines, the [\ion{O}{iii}]$\lambda$5007 line flux should be very week due to the low equilibrium temperature of the ionized gas owing to significant metal cooling. In the result, we found that our spectra show moderate EW of the [\ion{O}{iii}]$\lambda$5007 line similar to general quasars. This indicates nitrogen-loud quasars do not have extremely metal-rich gas clouds in NLRs. This suggests that strong nitrogen lines from broad-line regions (BLRs) are originated by exceptionally high abundances of nitrogen relative to oxygen without very high BLR metallicities. This result indicates that broad-emission lines of nitrogen are not good indicators of the BLR metallicity in some cases. On the other hand, we also investigate virial black-hole masses and Eddington ratios by using the H$\beta$ and \ion{C}{iv}$\lambda$1549 lines for our sample. As a result, we found that black-hole masses and Eddington ratios of nitrogen-loud quasars tend to be low and high relative to normal quasars, suggesting that nitrogen-loud quasars seem to be in a rapidly-accreting phase. This can be explained in terms of a positive correlation between Eddington ratios and nitrogen abundances of quasars, that is probably caused by the connection between the mass accretion onto black holes and nuclear star formation.}


\maketitle

\section{Introduction}\label{intr} 

Chemical evolution is one of the most crucial topics in extragalactic research since metal formations and its enrichments have close connections with the star formation in galaxies, corresponding to the galaxy evolution. The simplest way to investigate the chemical evolution of galaxies is by measuring metallicities or chemical abundances of galaxies at various redshifts and exploring its redshift evolution. Observationally, the gas-phase metallicity of star-forming galaxies in the local universe can be measured by using optical emission-line spectra, e.g., [\ion{O}{ii}]$\lambda$3727, H$\beta$, [\ion{O}{iii}]$\lambda$4959, [\ion{O}{iii}]$\lambda$5007, H$\alpha$, and [\ion{N}{ii}$]\lambda$6584 \citep[see][and references therein]{1979MNRAS.189...95P,2002ApJS..142...35K,2006A&A...459...85N,2016ApJ...828...67L}, and even at $z > 1$ current near-infrared (IR) observations can cover these lines \citep[e.g.,][]{2008A&A...488..463M,2016ApJ...827...74W}, with some caveats such as systematic errors in the calibration \citep[see, e.g.,][]{2008ApJ...681.1183K}. However, at $z > 3$, these diagnostics are more difficult because these emission lines are redshifted out of near-IR wavelength.

An alternative way to measuring metallicities of galaxies in the early universe is by focusing on the active galactic nucleus (AGN). In fact, this approach has mainly two advantages in the high-$z$ research as follows. First, thanks to its huge luminosity we can obtain high quality spectra even at high redshift. Second, because of various emission lines shown in rest-frame ultraviolet (UV) wavelength we can extract chemical properties from observed-frame optical spectra. The prominent emission-line spectra of AGNs provide chemical properties of gas clouds illuminated by the central engine: it is possible to investigate chemical properties even in the galactic scale, if we focus on emission lines emitted from the narrow-line regions, i.e., NLRs \citep[$R_{\rm NLR} \sim 10^{2-4}$ pc; e.g.,][]{2006A&A...456..953B,2006A&A...459...55B}, instead of broad-line regions (BLRs). Thus by using AGNs, we can reveal the chemical evolution in the early universe, closely related to the formation and evolution of galaxies. In addition, we emphasise that only AGNs allow us to estimate black-hole masses in the high-$z$ universe, based on the kinematical state of BLR clouds. Therefore, AGN spectra are useful also to understand the co-evolution of galaxies and supermassive black holes (SMBHs) in the high-$z$ epoch in addition to the chemical evolution.

\begin{table*} 
\caption{Journal of spectroscopic observations}
\label{jobs}
\centering
\begin{tabular}{lcccccr}
\hline\hline
\multicolumn{1}{c}{Target} & RA & Dec & $z^a$ & $i^b$ & Instrument & \multicolumn{1}{c}{Date}\\
\hline
SDSS J0357$-$0528     & 03 57 22.94 & $-$05 28 37.1 & 2.2887 & 19.09 & Subaru/IRCS & 23 Jan. 2012\\
SDSS J0931$+$3439     & 09 31 52.76 & $+$34 39 20.6 & 2.3051 & 19.44 & Subaru/IRCS & 23 Jan. 2012\\
SDSS J1211$+$0449$^c$ & 12 11 13.56 & $+$04 49 46.2 & 2.3184 & 18.87 & Subaru/IRCS & 23 Jan. 2012\\
SDSS J0756$+$2054     & 07 56 27.70 & $+$20 54 13.9 & 2.2109 & 19.11 & NTT/SOFI &  4 Feb. 2012\\
SDSS J0803$+$1742     & 08 03 25.04 & $+$17 42 39.0 & 2.0924 & 18.25 & NTT/SOFI &  3 Feb. 2012\\
SDSS J0857$+$2636     & 08 57 50.12 & $+$26 36 34.3 & 2.1394 & 18.09 & NTT/SOFI &  6 Feb. 2012\\
SDSS J0938$+$0900     & 09 38 01.18 & $+$09 00 04.4 & 2.1593 & 18.05 & NTT/SOFI &  5 Feb. 2012\\
SDSS J1036$+$1247     & 10 36 38.14 & $+$12 47 45.9 & 2.1557 & 18.66 & NTT/SOFI &  5 Feb. 2012\\
SDSS J1110$+$0456     & 11 10 48.93 & $+$04 56 08.0 & 2.2077 & 18.82 & NTT/SOFI &  3 Feb. 2012\\
SDSS J1130$+$1512     & 11 30 36.42 & $+$15 12 53.1 & 2.2271 & 18.56 & NTT/SOFI &  5 Feb. 2012\\
SDSS J1134$+$1500     & 11 34 23.00 & $+$15 00 59.2 & 2.1078 & 18.68 & NTT/SOFI &  6 Feb. 2012\\
SDSS J1308$+$1136     & 13 08 11.96 & $+$11 36 09.2 & 2.1012 & 17.70 & NTT/SOFI &  5 Feb. 2012\\
\hline
\end{tabular}
\tablefoot{$^a$ Redshift from \citet{2008ApJ...680..169S}; $^b$ SDSS $i$-band magnitude corrected for Galactic extinction in \citet{2008ApJ...679..962J}; $^c$ observed only with $H$ band.}
\end{table*} 

As regards diagnostics of AGN metallicities, we often recognize that nitrogen is one of the most useful elements, since it is a secondary element and its abundance is expected to be proportional to the square of the metallicity, i.e., N/H $\propto$ (O/H)$^2$. AGN spectra basically show some broad emission lines of nitrogen, e.g., \ion{N}{v}$\lambda$1240, \ion{N}{iv}]$\lambda$1486, and \ion{N}{iii}]$\lambda$1749, in the UV wavelength. \citet{1992ApJ...391L..53H,1993ApJ...418...11H} used emission-line flux ratios of \ion{N}{v}$\lambda$1240/\ion{C}{iv}$\lambda$1549 and \ion{N}{v}$\lambda$1240/\ion{He}{ii}$\lambda$1640 as indicators of the BLR metallicity \citep[see also][]{1999ARA&A..37..487H,2002ApJ...564..592H}. Subsequently, many papers follow this method based on nitrogen lines to investigate BLR metallicities \citep[e.g.,][]{2004ApJ...608..136W,2006A&A...447..157N,2009ApJ...696.1998D,2011A&A...527A.100M,2013ApJ...763...58S}.

In previous studies, emission-line flux ratios including nitrogen lines have been mostly used to estimate the BLR metallicity, $Z_{\rm BLR}$, as described above. Based on this method, it has been found that the BLR in most luminous quasars show super-solar metallicities \citep[e.g.,][]{1992ApJ...391L..53H,2003ApJ...589..722D,2006A&A...447..157N}, even for quasars at $z > 6$ \citep[e.g.,][]{2007ApJ...669...32K,2007AJ....134.1150J,2009A&A...494L..25J,2011Natur.474..616M}. Moreover, \citet{1980ApJS...42..333O} discovered an unusual quasar, Q0353$-$383, at $z = 1.96$ that has prominent broad emission lines of \ion{N}{iii}]$\lambda$1749, \ion{N}{iv}]$\lambda$1486, and \ion{N}{v}$\lambda$1240: this is called the nitrogen-loud quasar (hereafter N-loud quasar). \citet{2003ApJ...583..649B} claimed that these unusual nitrogen emissions in Q0353$-$383 is likely due to high metallicity in the BLR; the metallicity estimated by these nitrogen lines reaches $\sim 15 Z_\odot$ \citep[see also][]{1980ApJ...237..666O} if we assume the relation of N/H $\propto$ $Z_{\rm BLR}^2$. However, note that such extremely high metallicity is very hard to be reconciled with galaxy chemical evolutionary models. Therefore, the N-loud quasars raise a question whether the broad nitrogen lines are really reliable for estimating $Z_{\rm BLR}$, or not.

Motivated by this question, the rest-frame UV spectrum of N-loud quasars has been investigated so far. By using 293 N-loud quasars selected from the Sloan Digital Sky Survey \citep[SDSS;][]{2000AJ....120.1579Y}, \citet{2008ApJ...679..962J} reported that N-loud quasars may not have such high BLR metallicities, but just have unusually high relative abundance of nitrogen, because their rest-frame UV spectra are not significantly different from those of typical normal quasars, except for the nitrogen lines: this means that metallicity-sensitive emission line flux ratios excluding nitrogen do not show such extremely high metallicity expected from nitrogen lines. This explanation throws a doubt on the use of nitrogen lines for BLR metallicity measurements for quasars, at least in some cases. On the other hand, \citet{2014MNRAS.439..771B} aimed to clarify this controversial point by estimating BLR metallicities of N-loud quasars, and claimed that strong nitrogen lines of N-loud quasars are caused by high BLR metallicities instead of the high relative abundance of nitrogen. Note that, however, since this study was concluded only by using emission line flux ratios including nitrogen lines, i.e., \ion{N}{v}$\lambda$1240, \ion{N}{iv}]$\lambda$1486, and \ion{N}{iii}]$\lambda$1750, it would be still unclear whether N-loud quasars have extremely metal-rich gas clouds in BLRs or not: we need to estimate BLR metallicities by using emission line flux ratios including nitrogen lines and also using flux ratios excluding nitrogen lines, and compare them with each other.

To solve this problem regarding the origin of strong nitrogen lines seen in quasar spectra, we focus on NLR metallicities instead of BLR ones. If BLR metallicities of N-loud quasars are extremely high, NLR ones would be also relatively high although both absolute values are not the same \citep[e.g.,][]{2007ApJ...664L..75F,2014MNRAS.438.2828D}. Recently, \citet{2012A&A...543A.143A} investigated the NLR metallicity of a N-loud quasar at $z = 3.16$, SDSS J1707$+$6443, by using the [\ion{O}{iii}]$\lambda$5007 emission line shown in rest-frame optical wavelengths: if the NLR metallicity is remarkably high as suggested by strong UV nitrogen lines, the [\ion{O}{iii}]$\lambda$5007 line flux should be very weak due to the low equilibrium temperature of the ionized gas owing to significant metal cooling. We also examined its other parameters (e.g., ionization parameter and hydrogen density of NLR gas clouds) by using fluxes of [\ion{Ne}{v}]$\lambda$3426, [\ion{O}{ii}]$\lambda$3727, and [\ion{Ne}{iii}]$\lambda$3869, which are useful to estimate correctly the NLR metallicity. In the result, we found that the NLR in this N-loud quasar is not extremely metal rich, that is not consistent with the naive expectation from unusually strong nitrogen broad lines. Note that, however, it is difficult to conclude that the N-loud quasar is not extremely metal rich \emph{generally} because this result was obtained only from one object. Therefore, additional data set of rest-frame optical spectra of N-loud quasars are crucially needed to investigate whether or not BLR metallicities of N-loud quasars are extremely high universally.

Furthermore, if N-loud quasars do not have extremely high metallicities as \citet{2012A&A...543A.143A} claimed, unusually high nitrogen abundance relative to oxygen is required. Since the chemical properties of gas in galactic nuclei are closely related to the past star-formation history and the gas radial motion, we focus on the relation between the mass accretion to SMBHs and the nuclear star formation. Recently, by using optical spectra of 2383 SDSS quasars at $2.3 < z < 3.0$, \citet{2011A&A...527A.100M} investigated the dependence of the emission-line flux ratios (BLR metallicity indicators) on the Eddington ratio. In the result, they found that the Eddington ratio depends on the emission-line flux ratios involving \ion{N}{v}$\lambda$1240, while it does not correlate with the other emission-line flux ratios without the nitrogen line \citep[see Figure~6 in][]{2011A&A...527A.100M}. This result means that the relative nitrogen abundance seems to depend on the Eddington ratio. This indicates that a rapid mass accretion onto SMBHs is associated with a post-starburst phase, when asymptotic giant brach (AGB) stars enrich the interstellar medium with nitrogen. By considering this relation, we can obtain one possible scenario that N-loud quasars are in an AGN active phase, i.e., a high accretion phase. This would be helpful to understand the connection between AGN and star formation activities, closely related to the co-evolution of galaxies and SMBHs. In this sense, therefore, it is also important to examine whether N-loud quasars show high Eddington ratios or not.

\begin{figure*} 
\centering
\includegraphics[width=8.5cm]{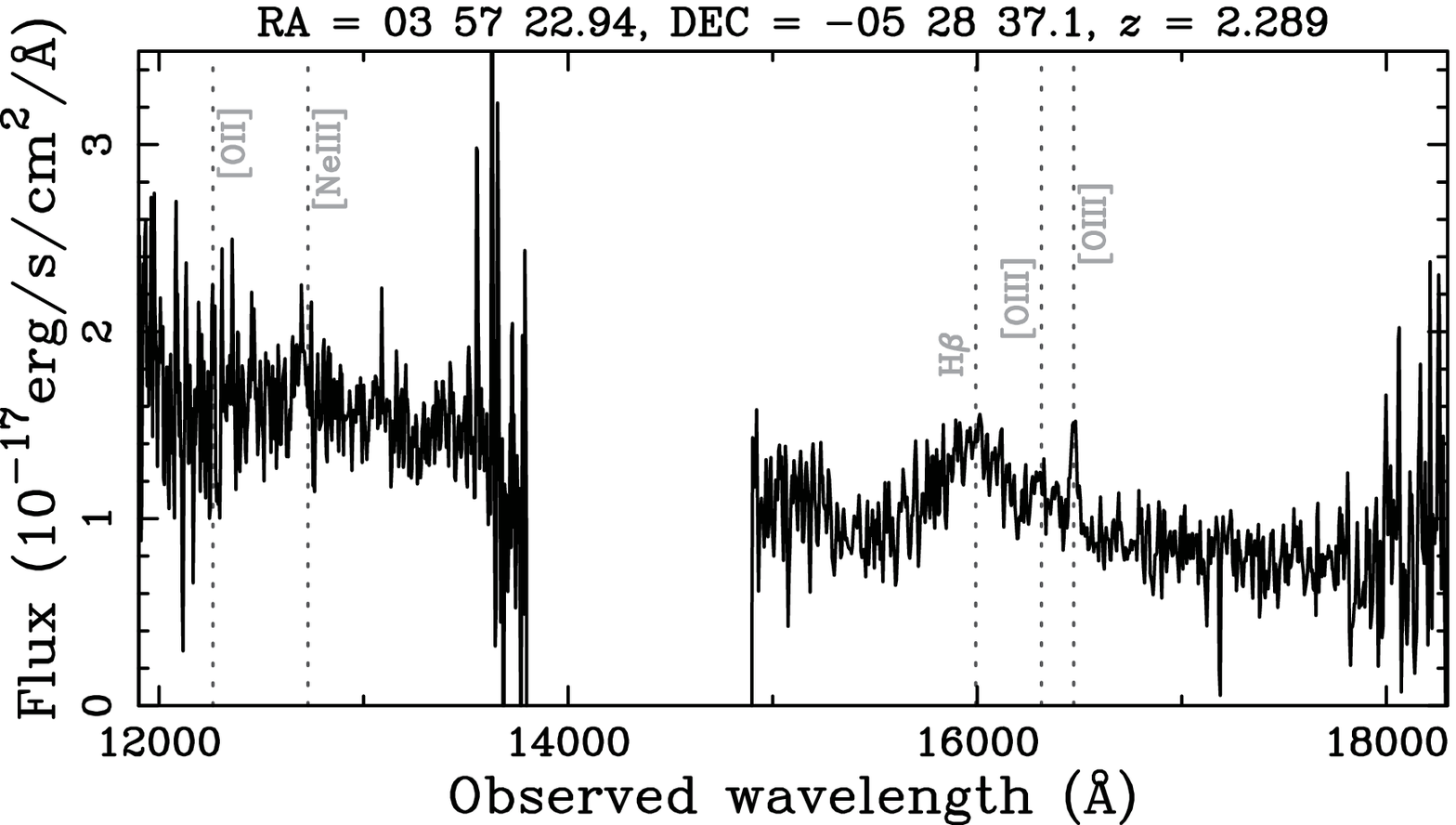}
\hspace{0.2cm}
\includegraphics[width=8.5cm]{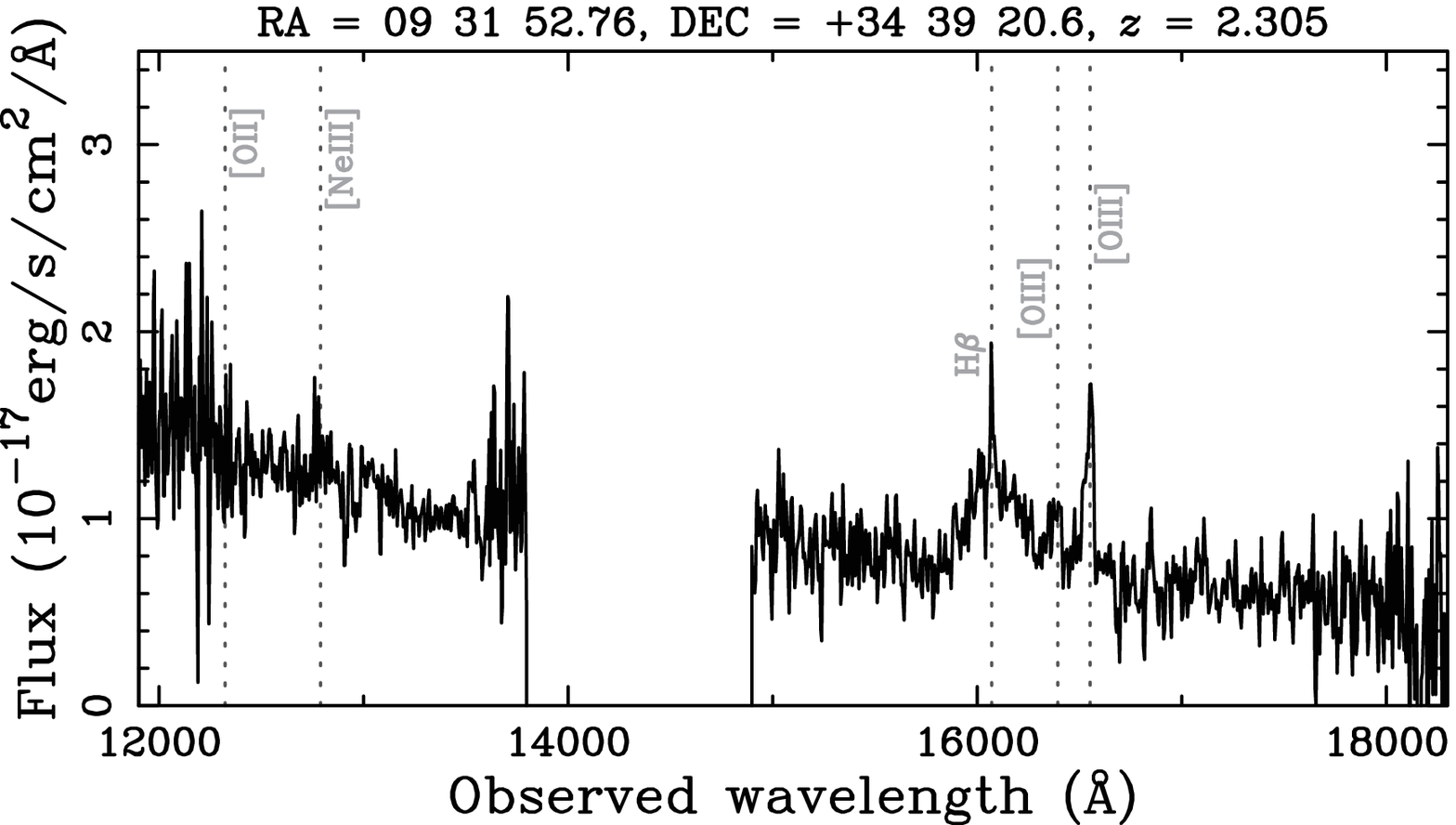}

\vspace{0.2cm}
\includegraphics[width=8.5cm]{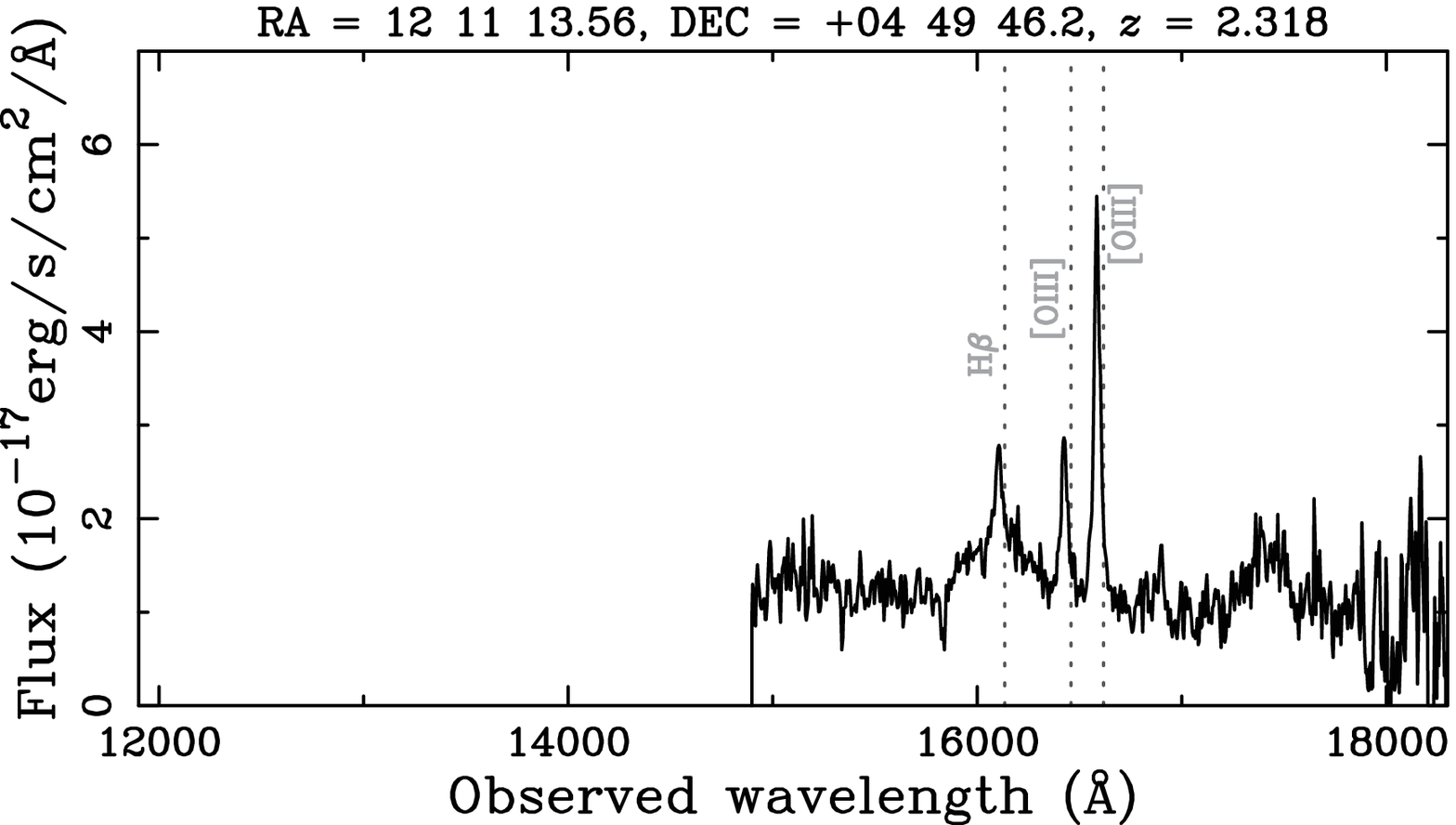}
\hspace{0.2cm}
\includegraphics[width=8.5cm]{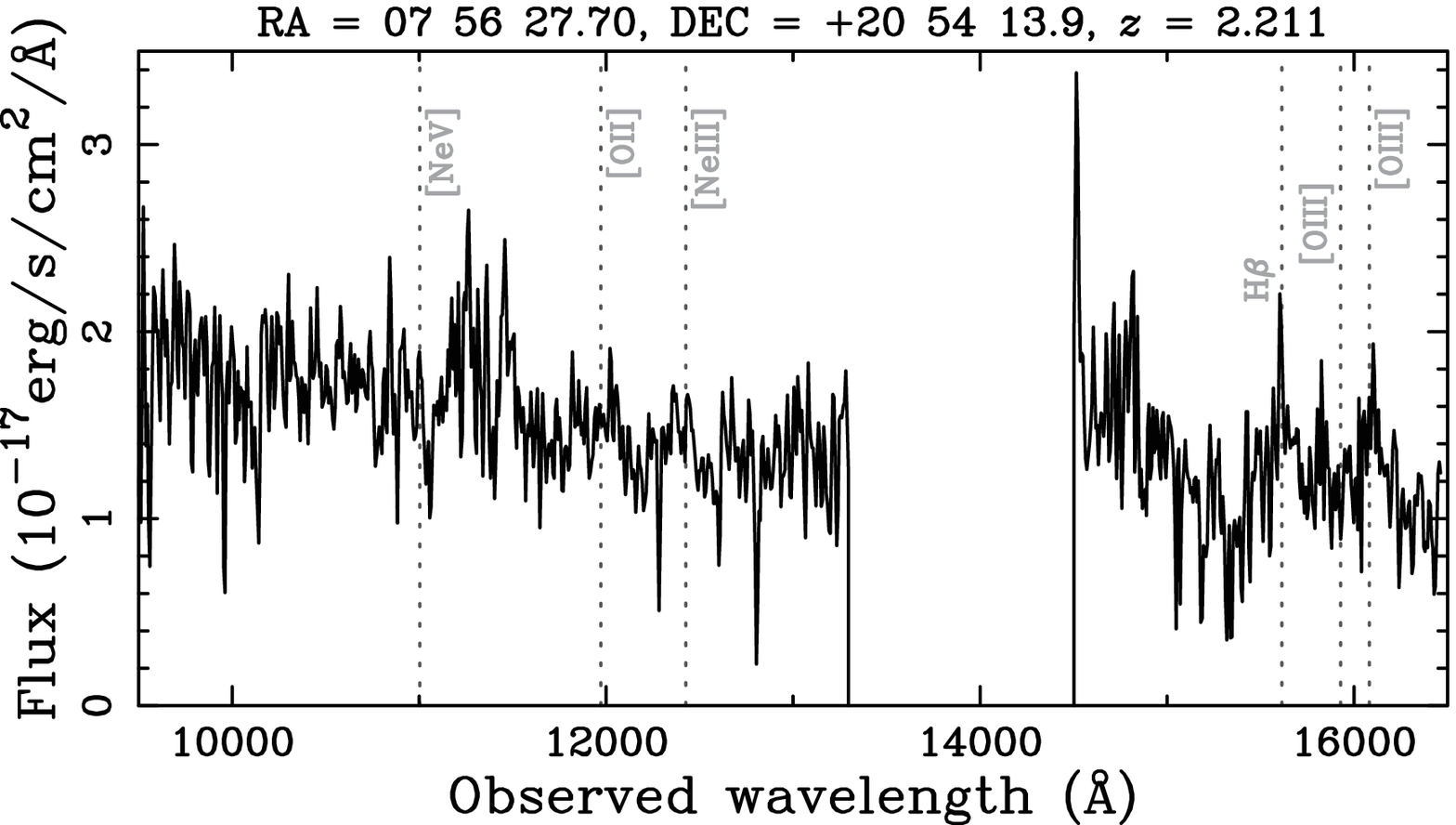}

\vspace{0.2cm}
\includegraphics[width=8.5cm]{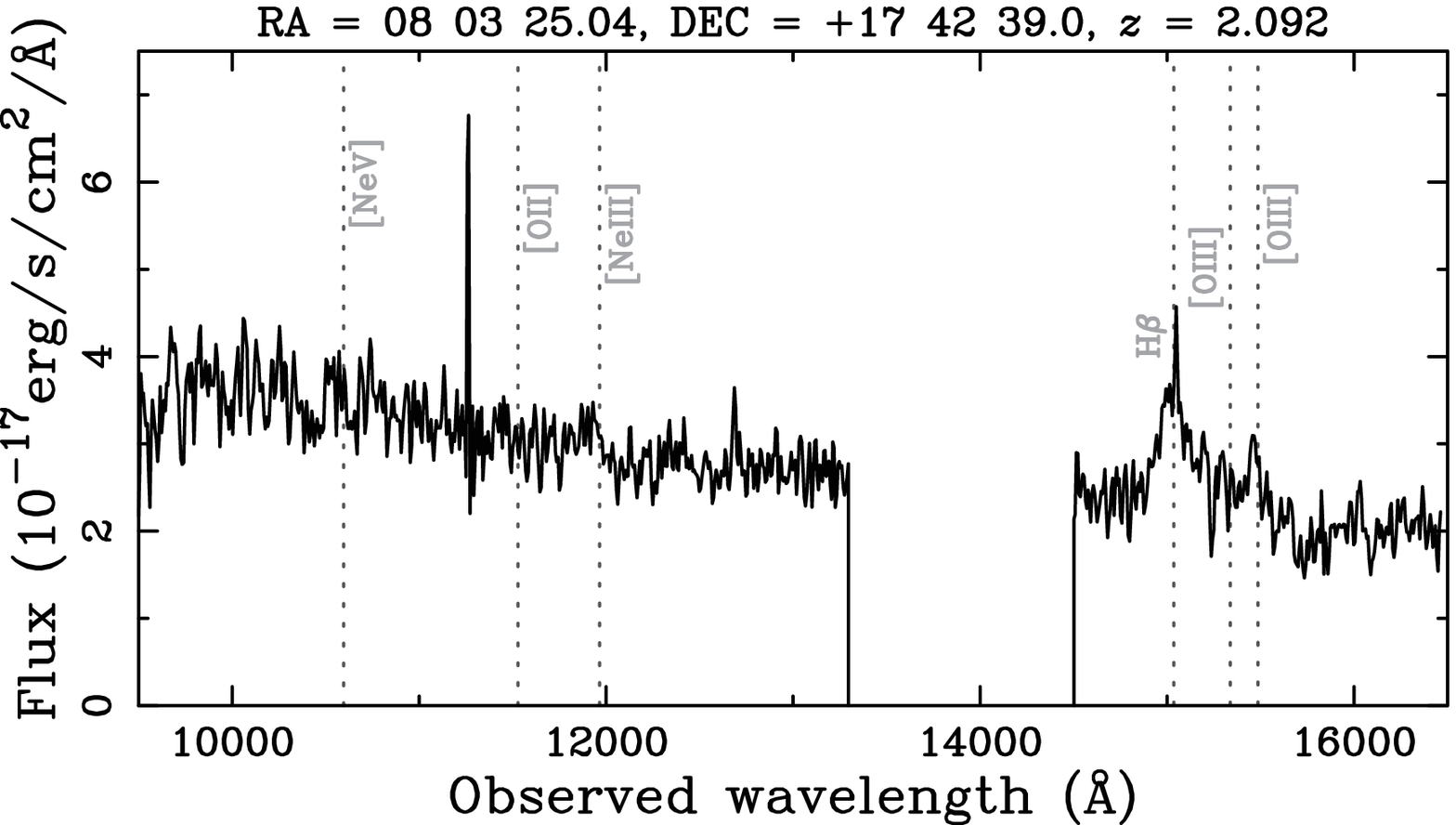}
\hspace{0.2cm}
\includegraphics[width=8.5cm]{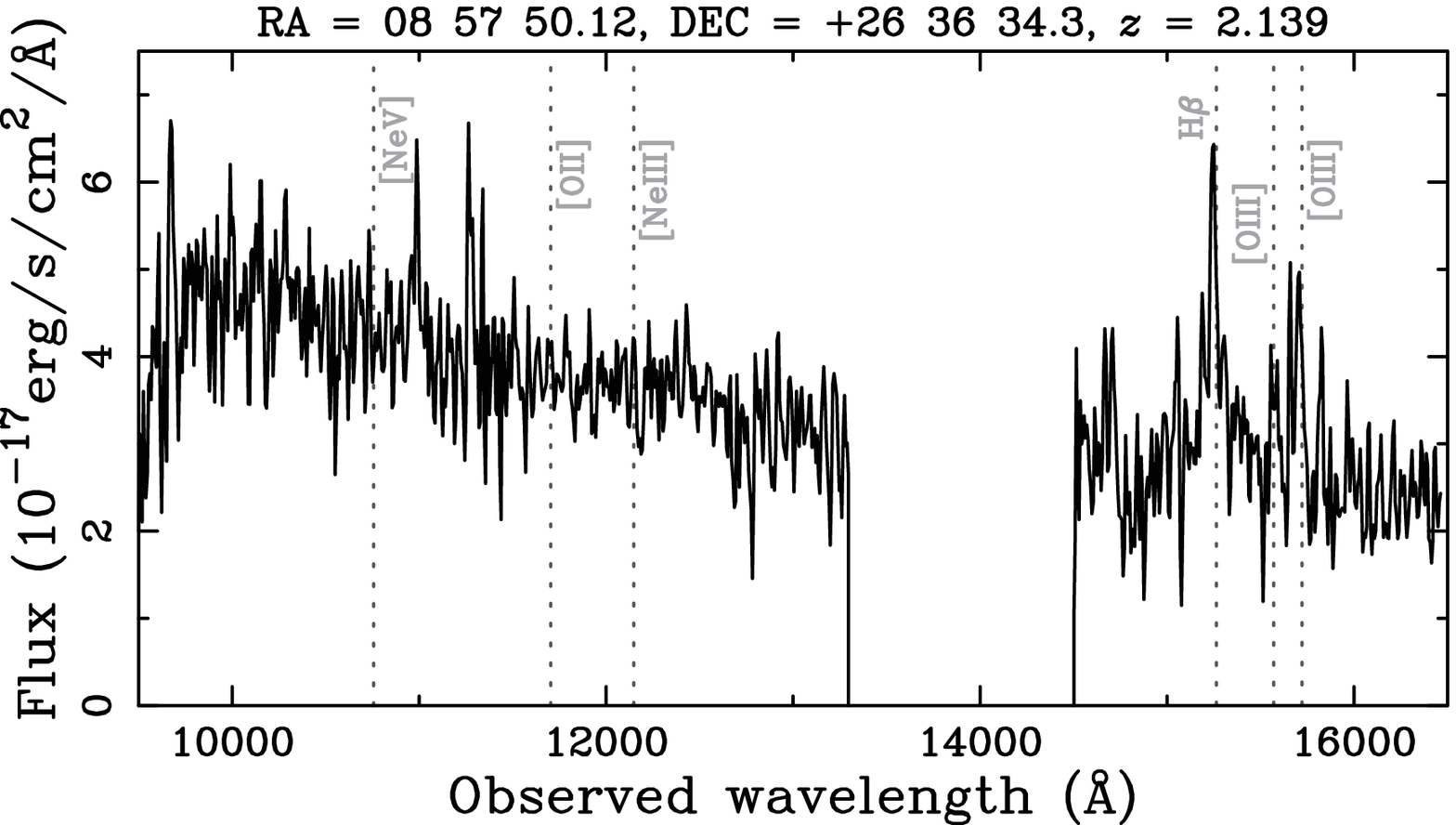}

\vspace{0.2cm}
\includegraphics[width=8.5cm]{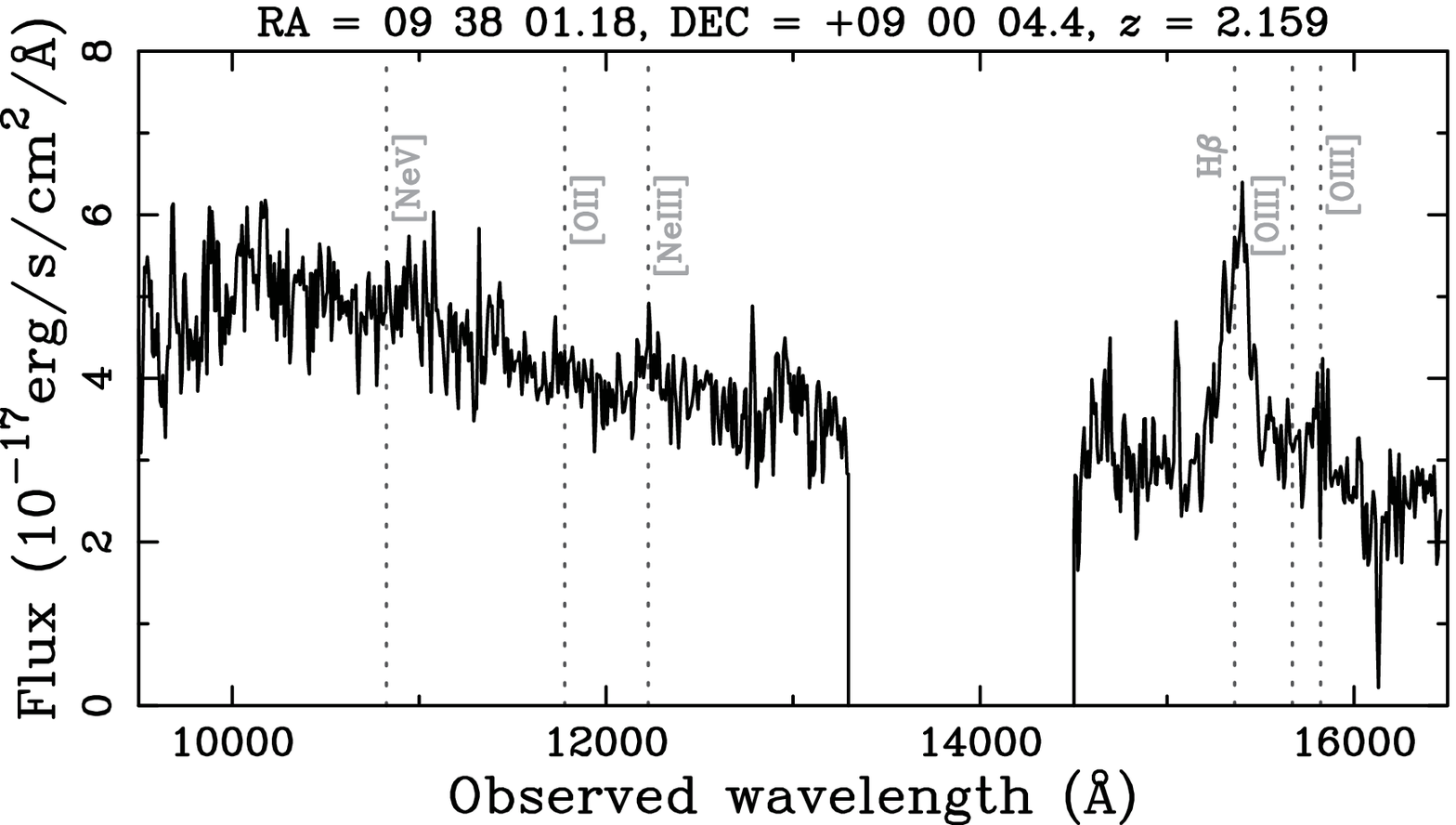}
\hspace{0.2cm}
\includegraphics[width=8.5cm]{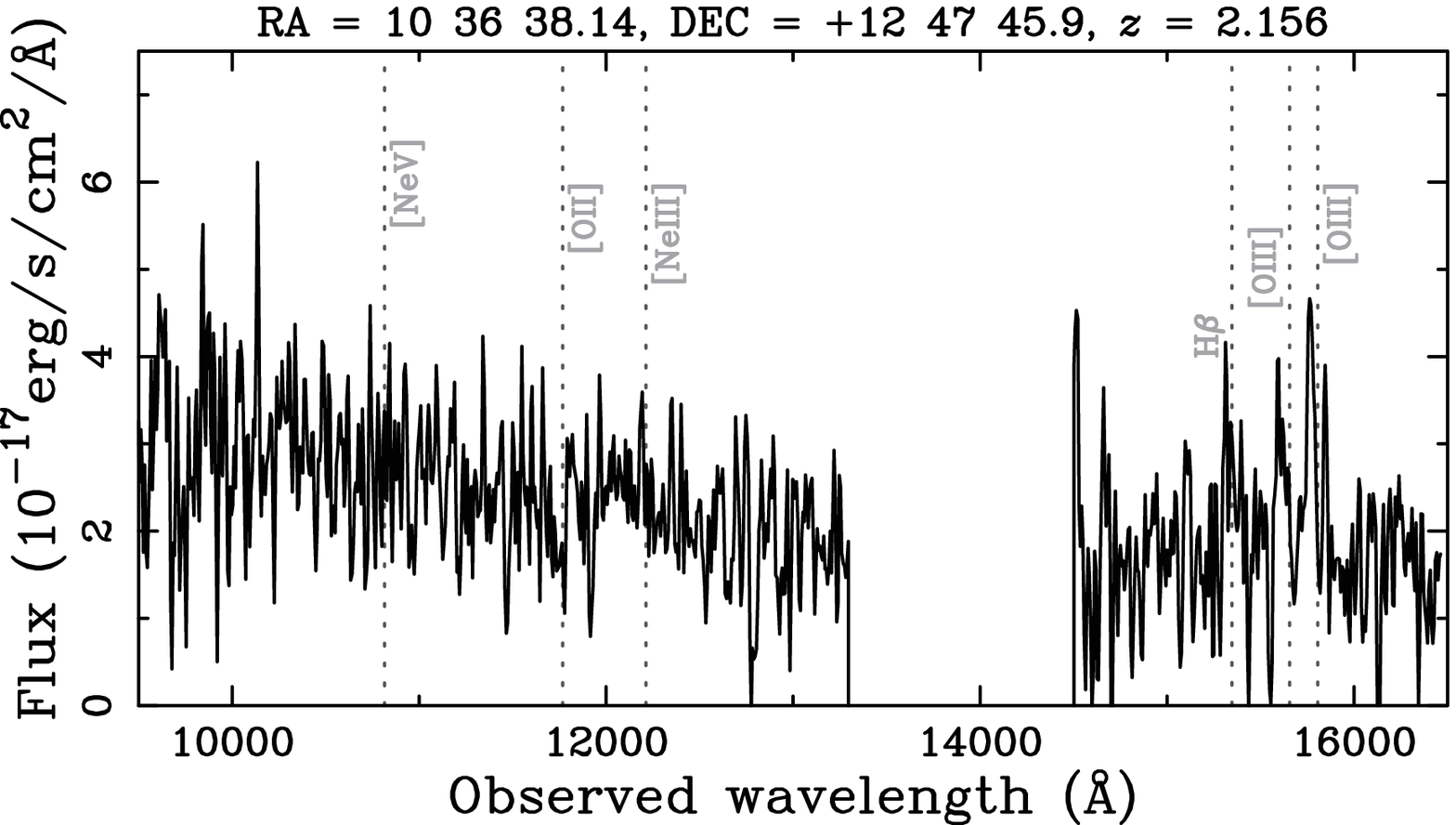}
\caption{Near-infrared spectra of N-loud quasars obtained in our IRCS and SOFI runs, adopting the two-pixel binning in the wavelength direction. Grey dotted lines show the wavelengths of some rest-frame optical emission lines, i.e., [\ion{Ne}{v}]$\lambda$3426, [\ion{O}{ii}]$\lambda$3727, [\ion{Ne}{iii}]$\lambda$3869, H$\beta$, [\ion{O}{iii}]$\lambda$4959, and [\ion{O}{iii}]$\lambda$5007 lines, expected from redshifts given in \citet{2008ApJ...680..169S}.}
\label{f:spec}
\end{figure*}

\setcounter{figure}{0}
\begin{figure*}
\centering
\includegraphics[width=8.5cm]{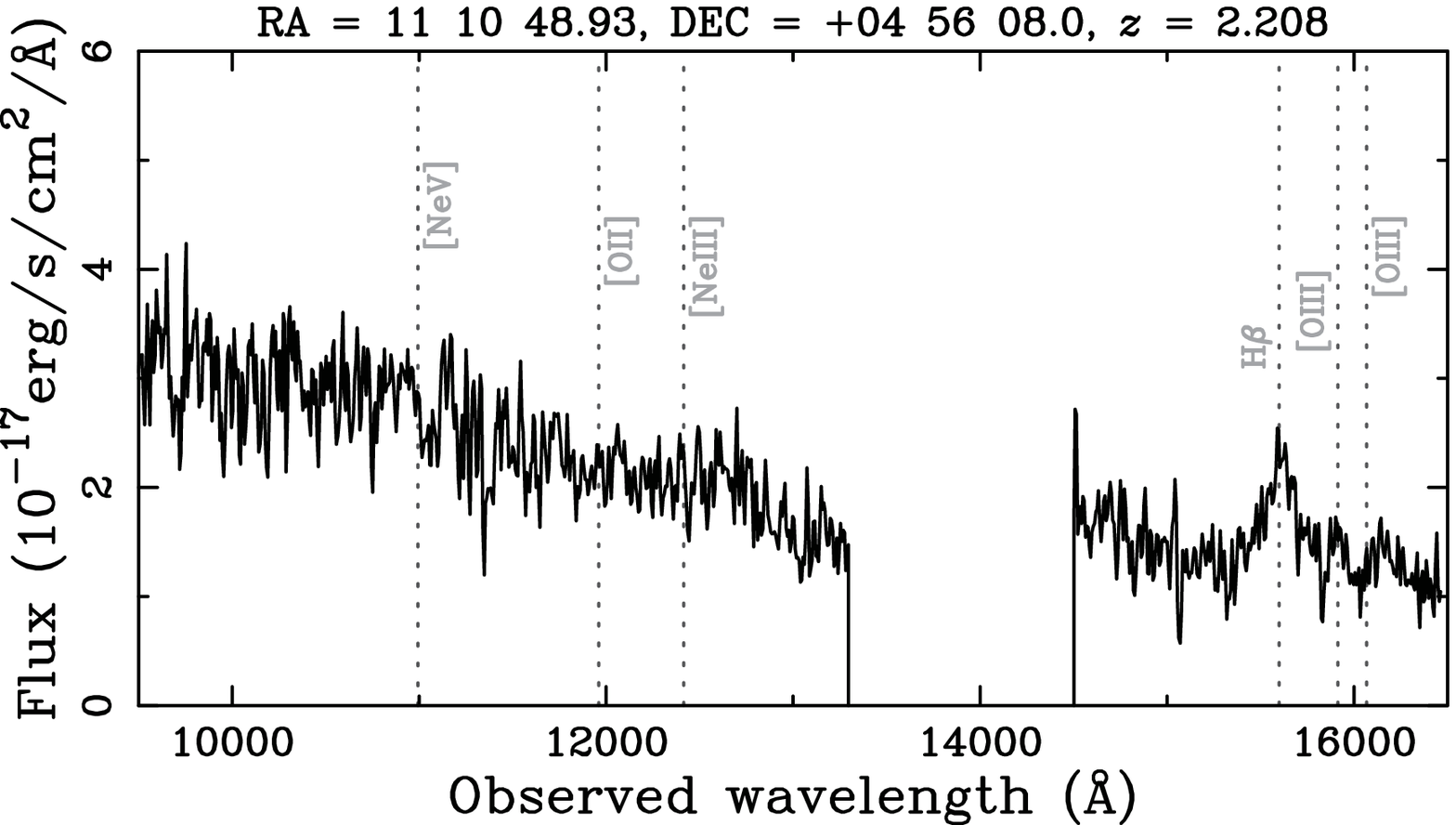}
\hspace{0.2cm}
\includegraphics[width=8.5cm]{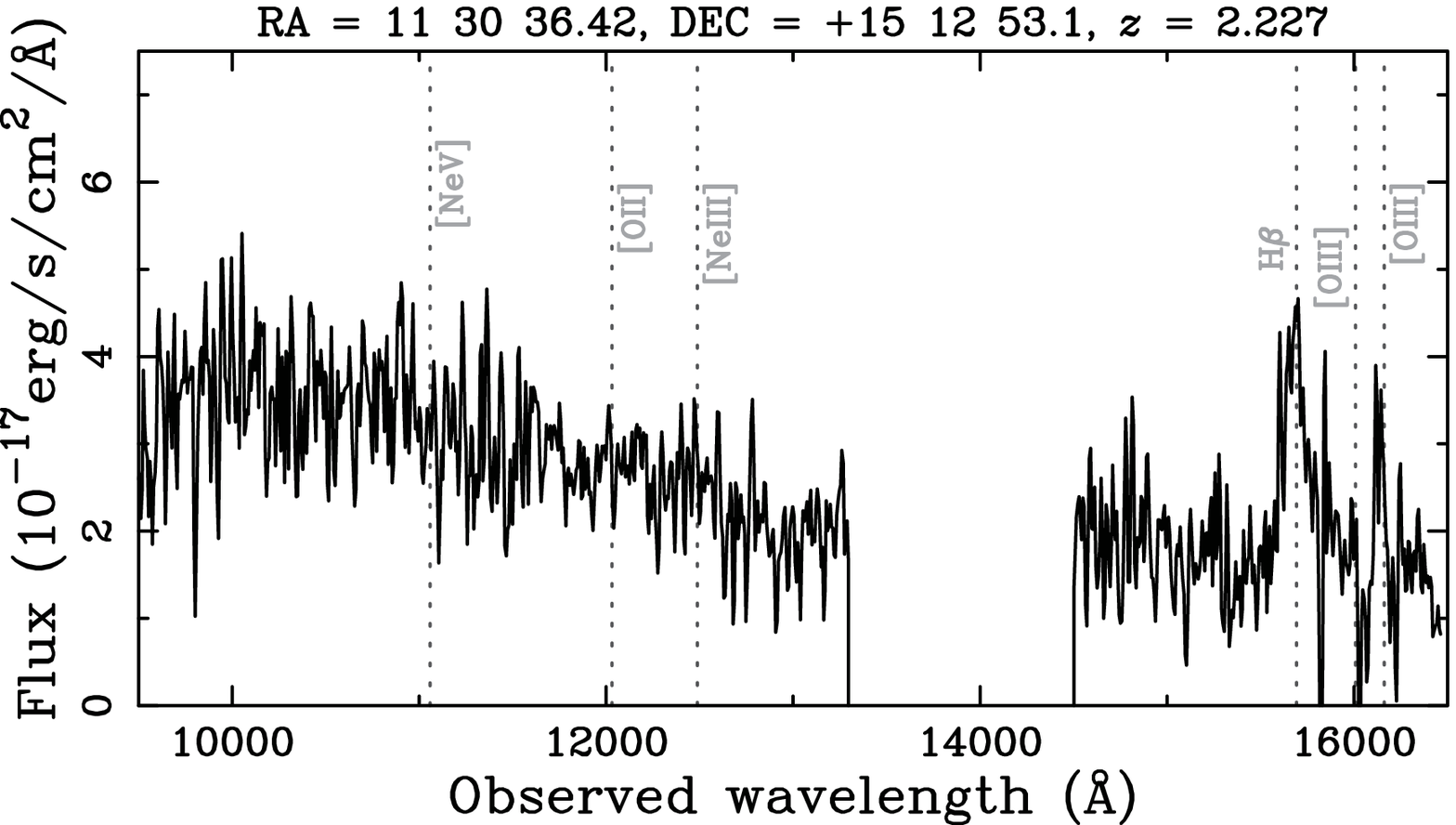}

\vspace{0.2cm}
\includegraphics[width=8.5cm]{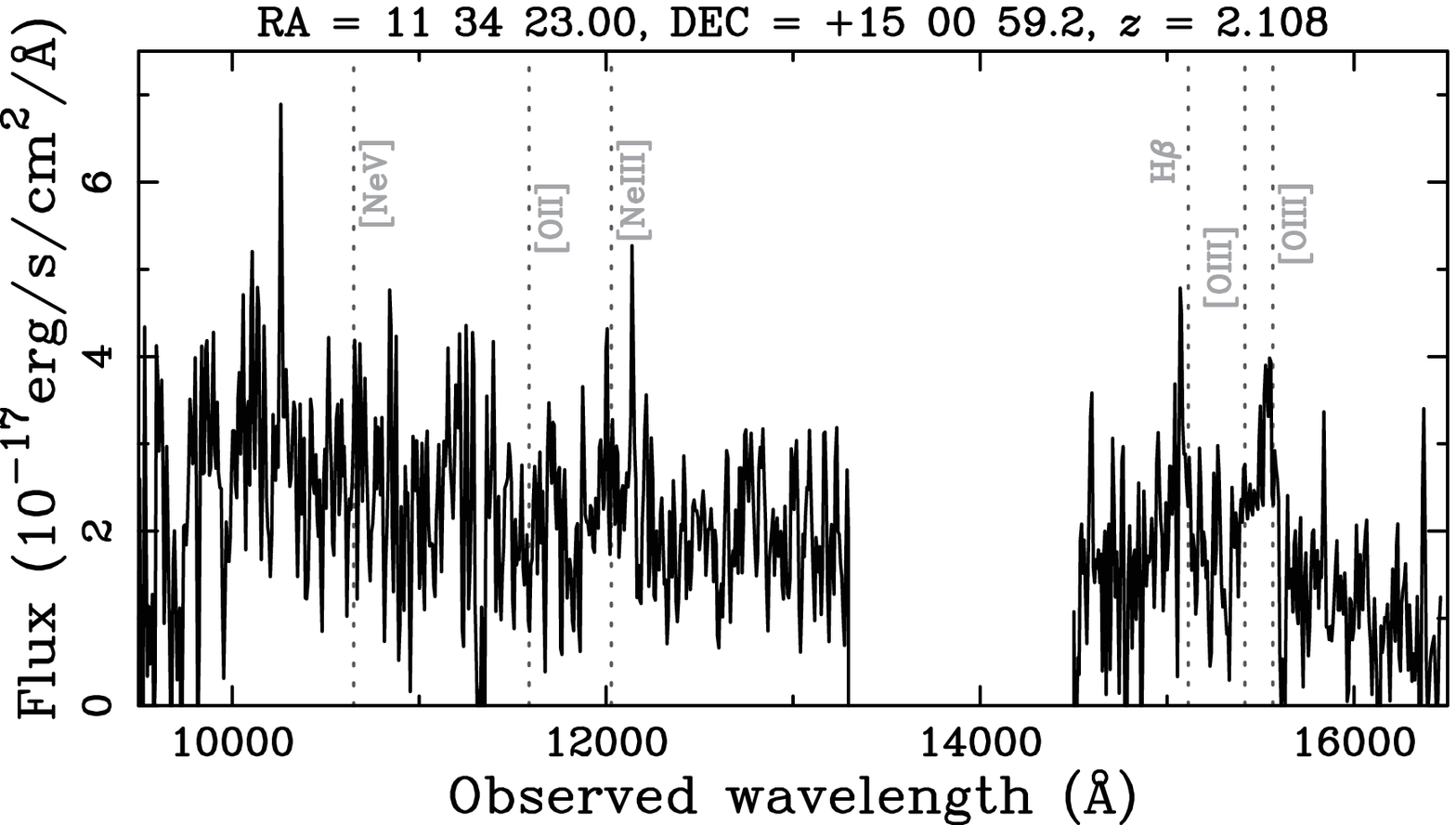}
\hspace{0.2cm}
\includegraphics[width=8.5cm]{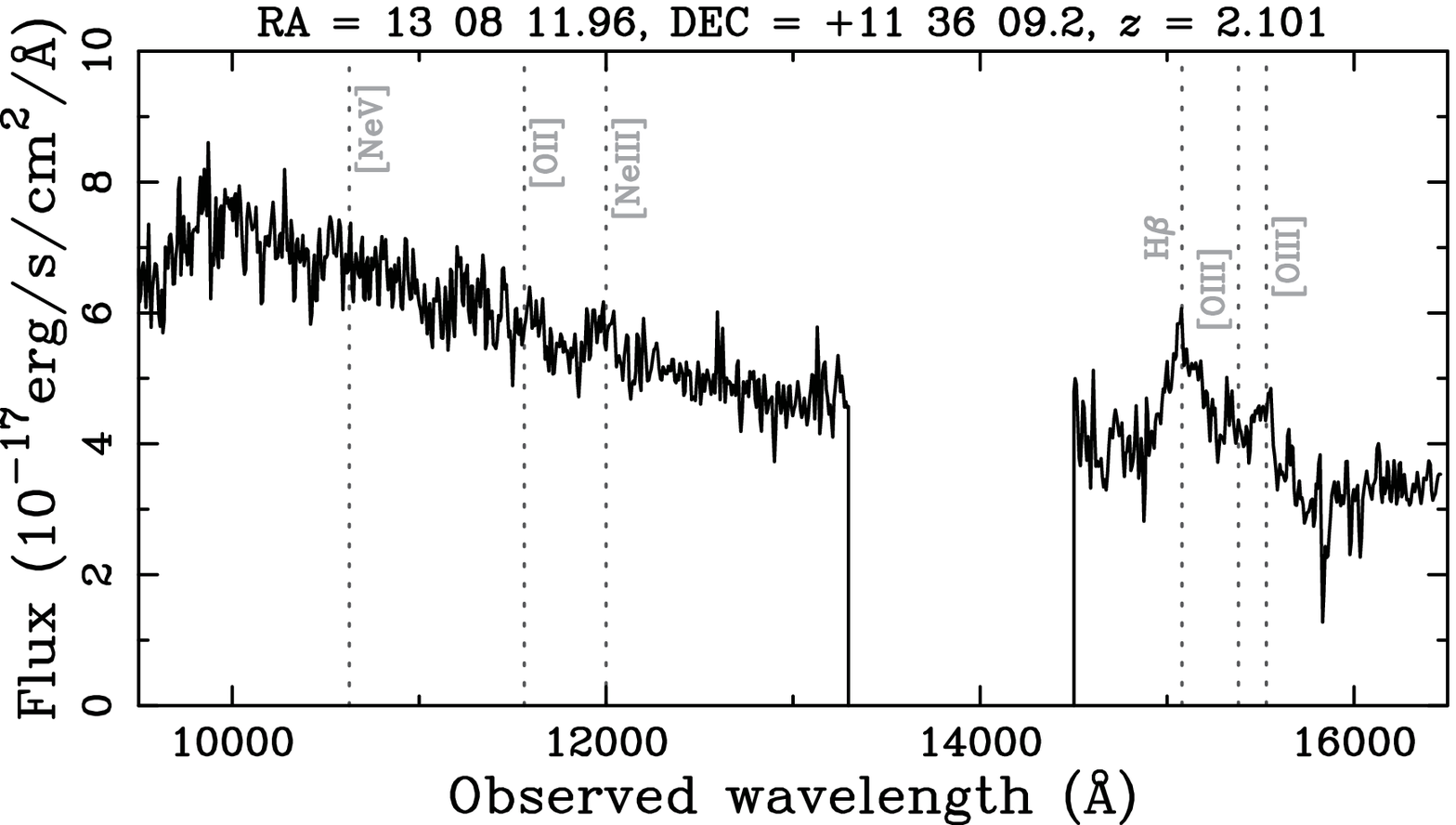}
\caption{Continued.}
\end{figure*} 

\begin{figure*} 
\centering
\includegraphics[width=8.5cm]{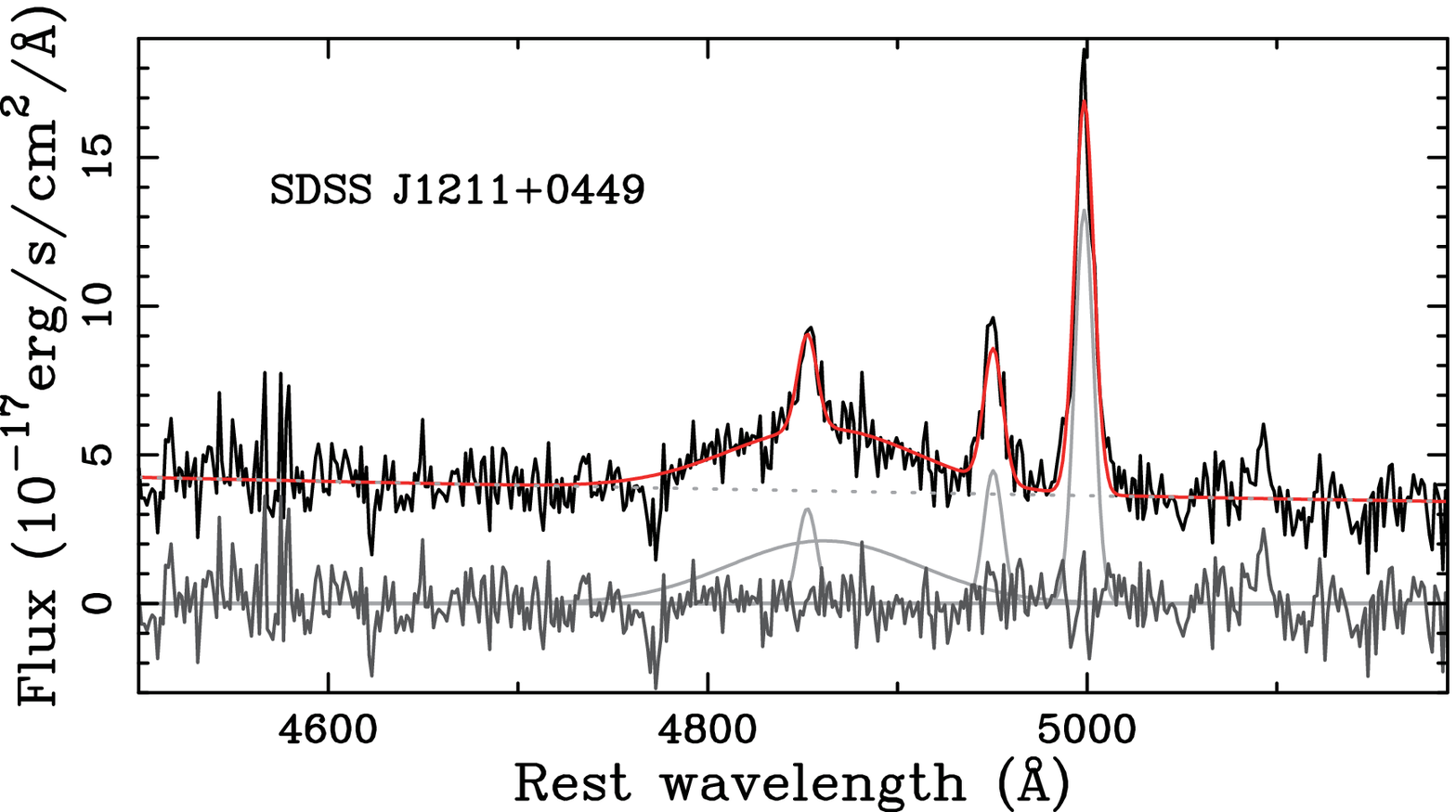}
\hspace{0.2cm}
\includegraphics[width=8.5cm]{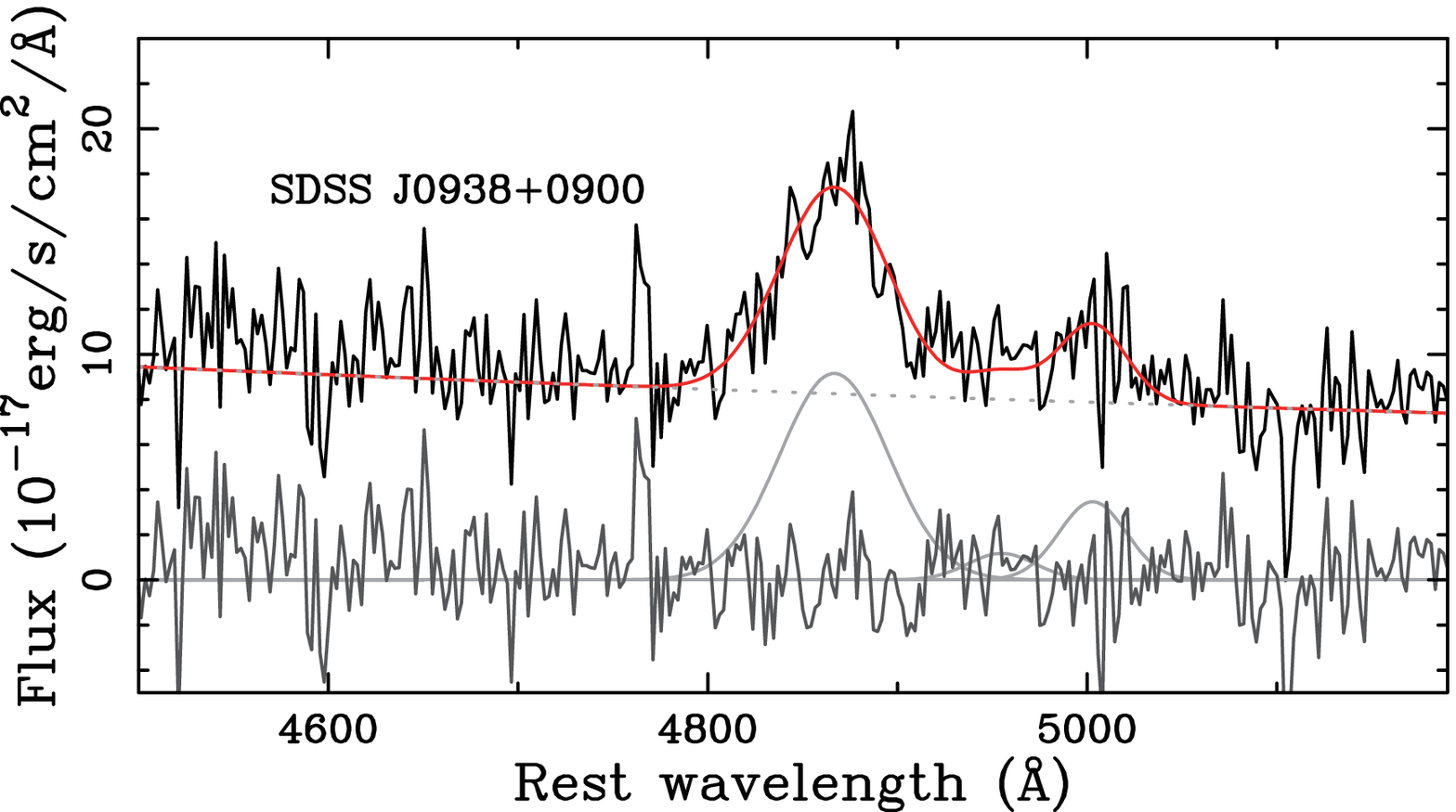}
\caption{Examples of spectral fittings in rest-frame optical spectra of two N-loud quasars (SDSS J1211$+$0449 and SDSS J0938$+$0900). Red lines show total fitting results. Grey-dotted lines denote fits of continuum and individual emission-line components are indicated as light-grey gaussian curves. Dark-grey lines denote the residuals.}
\label{f:lfit}
\end{figure*} 

In this paper, we present near-infrared spectra of 12 N-loud quasars and investigate the spectroscopic properties of them. Specifically, we estimate NLR metallicities, black-hole masses, and Eddington ratios by using emission lines in the rest-frame optical wavelength. In Section~\ref{obse}, we describe our sample of N-loud quasars and observations. Section~\ref{spec} shows the spectral measurements. We summarize our results in Section~\ref{resu} and discuss the interpretation for results in Section~\ref{disc}. The summary is given in Section~\ref{conc}. Here, we adopt a concordance cosmology with ($\Omega_{\rm M}, \Omega_\Lambda$) = (0.3, 0.7) and $H_0 = 70$ km s$^{-1}$ Mpc$^{-1}$.

\section{Observations and data reduction}\label{obse} 

\subsection{Target selection} 

We select our targets from the N-loud quasar catalogue of \cite{2008ApJ...679..962J}. They collected 293 N-loud quasars from the SDSS Fifth Data Release \citep{2007AJ....134..102S} with following criteria: (1) rest-frame equivalent widths (EW) of \ion{N}{iv}]$\lambda$1486 or \ion{N}{iii}]$\lambda$1750 $> 3${\AA} which is significantly higher than those of normal quasars, i.e., $\sim 0.4${\AA} for \ion{N}{iii}]$\lambda$1750 line and no detection for \ion{N}{iv}]$\lambda$1486 line typically \citep[see also][]{2001AJ....122..549V}; (2) the redshift range $1.7 < z < 4.0$ covering one or two broad nitrogen lines in SDSS optical spectra; and (3) $i' < 20.1$. To obtain emission lines of [\ion{O}{iii}]$\lambda$5007, H$\beta$, [\ion{Ne}{iii}]$\lambda$3869, [\ion{O}{ii}]$\lambda$3727, and [\ion{Ne}{v}]$\lambda$3426 in near-infrared wavelength range, in this study we select 12 quasars at $2.09 < z < 2.57$ from the catalogue with criteria of observable coordinates and apparent magnitudes, given in Table~\ref{jobs}.

\begin{table*} 
\caption{Results of spectral fitting.}
\label{t:resu}
\centering
\begin{tabular}{lcccccr}
\hline\hline
\multicolumn{1}{c}{Object} & $F_{\rm [\ion{O}{iii}]}$ & $\log$EW$_{\rm [\ion{O}{iii}]}$$^a$ & $\log$FWHM$_{\rm H\beta}$$^b$ & $\log (\lambda L_{\rm 5100})$ & $\log M_{\rm BH,H\beta}$$^c$ & \multicolumn{1}{c}{$\log (L/L_{\rm Edd})_{\rm H\beta}$$^c$}\\
& [10$^{-17}$ ergs s$^{-1}$ cm$^{-2}$] & [\AA] & [km s$^{-1}$] & \multicolumn{1}{c}{[ergs s$^{-1}$]} & [$M_\odot$] &\\
\hline
SDSS J0357$-$0528 &  28.19$\pm$4.80  & 0.980$\pm$0.205 & 3.953$\pm$0.014 & 45.768$\pm$0.001 & 9.699$\pm$0.028 & $-$1.065$\pm$0.028\\
SDSS J0931$+$3439 &  48.77$\pm$3.30  & 1.338$\pm$0.106 & 3.788$\pm$0.014 & 45.641$\pm$0.007 & 9.306$\pm$0.028 & $-$0.799$\pm$0.029\\
SDSS J1211$+$0449 & 166.49$\pm$9.27  & 1.659$\pm$0.212 & 3.828$\pm$0.014 & 45.873$\pm$0.004 & 9.502$\pm$0.028 & $-$0.763$\pm$0.029\\
SDSS J0756$+$2054 &  74.31$\pm$14.89 & 1.360$\pm$0.226 & 3.604$\pm$0.050 & 45.775$\pm$0.000 & 9.006$\pm$0.100 & $-$0.364$\pm$0.100\\
SDSS J0803$+$1742 &  90.54$\pm$19.48 & 1.157$\pm$0.230 & 3.826$\pm$0.014 & 46.006$\pm$0.001 & 9.565$\pm$0.028 & $-$0.693$\pm$0.028\\
SDSS J0857$+$2636 & 106.50$\pm$21.39 & 1.139$\pm$0.295 & 3.733$\pm$0.059 & 46.116$\pm$0.003 & 9.434$\pm$0.119 & $-$0.451$\pm$0.119\\
SDSS J0938$+$0900 & 149.28$\pm$37.67 & 1.279$\pm$0.274 & 3.627$\pm$0.014 & 46.131$\pm$0.003 & 9.230$\pm$0.028 & $-$0.232$\pm$0.028\\
SDSS J1036$+$1247 & 176.64$\pm$21.00 & 1.583$\pm$0.328 & 3.208$\pm$0.041 & 45.899$\pm$0.007 & 8.275$\pm$0.081 &    0.490$\pm$0.082\\
SDSS J1110$+$0456 & $< 40.92^d$      & $< 0.195^d$     & 3.640$\pm$0.013 & 45.822$\pm$0.001 & 9.102$\pm$0.026 & $-$0.413$\pm$0.026\\
SDSS J1130$+$1512 & 103.62$\pm$26.12 & 1.361$\pm$0.387 & 3.441$\pm$0.009 & 45.921$\pm$0.001 & 8.752$\pm$0.018 &    0.035$\pm$0.018\\
SDSS J1134$+$1500 & 222.87$\pm$57.56 & 1.747$\pm$0.566 & 3.575$\pm$0.030 & 45.809$\pm$0.007 & 8.964$\pm$0.061 & $-$0.288$\pm$0.061\\
SDSS J1308$+$1136 & 165.45$\pm$30.19 & 1.204$\pm$0.193 & 3.858$\pm$0.019 & 46.220$\pm$0.002 & 9.737$\pm$0.037 & $-$0.650$\pm$0.037\\
\hline
\end{tabular}
\tablefoot{$^a$ Logarithmic rest-frame equivalent width of the [\ion{O}{iii}]$\lambda$5007 line; $^b$ logarithmic rest-frame line width of the H$\beta$ broad line after the correction for instrument broadening; $^c$ only statistical errors are given instead of systematic ones; $^d$ 3$\sigma$ upper limits given for the [\ion{O}{iii}]$\lambda$5007-undetected object.}
\end{table*} 

\subsection{Near-infrared spectroscopic observations} 

We observed our targets by using the Infrared Camera and Spectrograph \citep[IRCS;][]{2000SPIE.4008.1056K} with the adaptive optics system with 188 elements \citep[AO188;][]{2008SPIE.7015E..25H,2010SPIE.7736E..21H} on the Subaru Telescope, and with the Son of ISAAC \citep[SOFI;][]{1998Msngr..91....9M} on the New Technology Telescope (NTT). The observational logs are given in Table~\ref{jobs}.

Regarding the IRCS observation, we obtained $J$- and $H$-band spectra of three targets at $z \sim 2.3$ by using a 0\farcs225-width slit. Thanks to AO188 with the laser guide star (LGS) mode, we could achieve typical point-spread function sizes of $\sim$ 0.25\arcsec\ around 13000\AA\ and $\sim$ 0.21\arcsec\ at 16000\AA. In $H$-band observation, we aimed at measuring the emission lines of [\ion{O}{iii}]$\lambda$5007 and H$\beta$, whose expected wavelength is 16000{\AA} $< \lambda_{\rm obs} <$ 16600{\AA}, to investigate NLR metallicities and black-hole masses. On the other hand, in $J$-band observation we focus on emission lines of [\ion{O}{ii}]$\lambda$3727 and [\ion{Ne}{iii}]$\lambda$3869, redshifted to the range at 12200{\AA} $< \lambda_{\rm obs} <$ 12800\AA. These emission lines would be useful to investigate properties of NLR gases (e.g., ionization parameters and hydrogen densities of clouds). The spectral dispersion is $\sim 4.68${\AA} pixel$^{-1}$ and the spectral resolution estimated from OH emission lines was $\sim 1000$. The total exposure times for each target are roughly 7000 sec for $J$ band and 2500 sec for $H$ band. The unit exposure times were 300 -- 600 sec.

On the other hand, the SOFI observations were performed with Grism Blue Filter (GBF) to cover the wavelength at 9500{\AA} $< \lambda_{\rm obs} <$ 16400\AA, where we can cover five emission lines of remaining nine N-loud quasars at $2.1 \la z \la 2.2$ simultaneously. Typical seeing size is $\sim$ 1.5\arcsec. We used a long slit with 0\farcs6 width and the dispersion is $\sim 6.93${\AA} pixel$^{-1}$. The spectral resolution was $\sim 650$. Total exposure times are $\sim 7200$ sec for each target, adopting an unit exposure time of $\sim$ 600 sec.

\subsection{Data reduction} 

For all spectra, standard data-reduction procedures were performed using available IRAF\footnote{IRAF is distributed by the National Optical Astronomy Observatory, which is operated by the Association of Universities for Research in Astronomy (AURA) under a cooperative agreement with the National Science Foundation.} tasks. By using dome-flat frames, object frames were flat-fielded and thus the first sky subtraction was performed with the A$-$B method; we subtracted pairs of subsequent frames with the target at different positions along the slit. Cosmic-ray events were removed by using the {\tt lacos\_spec} task \citep{2001PASP..113.1420V}. We extracted the one-dimensional spectra with the {\tt apall} task, using eight pixel ($\sim$ 0.4\arcsec) aperture for IRCS and 11 pixels ($\sim$ 3.0\arcsec) for SOFI. In this extraction, we fit the residuals of sky background and subtracted them. The wavelength calibration was executed with arc lamp frames. Finally, flux calibration and telluric absorption correction were carried out by using observed spectra of telluric standard stars. We took defocused data for bright standard star observations to avoid the saturation of the detector: this step would affect the flux calibration, that obtained fluxes are overestimated due to the extensive slit loss of defocused data.

To correct observational uncertainties, e.g., the slit loss, mainly in observations of standard stars, and weather condition, we re-calibrated fluxes by using photometric magnitudes of the targets themselves. We obtained near-infrared magnitudes of $J$ and $H$ bands from $i$-band magnitude using a colour-$z$ relation of SDSS-UKIDSS quasars in \citet{2010MNRAS.406.1583W}, since there is almost no near-infrared magnitude of our targets. Note that only two objects (i.e., SDSS J0938$+$0900 and SDSS J1308$+$1136) have been observed with $J$ and $H$ bands through the United Kingdom Infrared Telescope (UKIRT) Infrared Deep Sky Survey \citep[UKIDSS;][]{2007MNRAS.379.1599L}, and we confirmed these observed magnitudes are consistent with estimated ones describe above in uncertainties of $\sim 2\%$. By comparing the extrapolated photometric magnitudes with spectroscopic magnitudes measured from our spectra, we scaled fluxes of the spectra; the mean value of the scaling factor is $\sim 0.8$. Final reduced spectra are shown in Figure~\ref{f:spec}.

\begin{figure} 
\centering
\includegraphics[width=8.5cm]{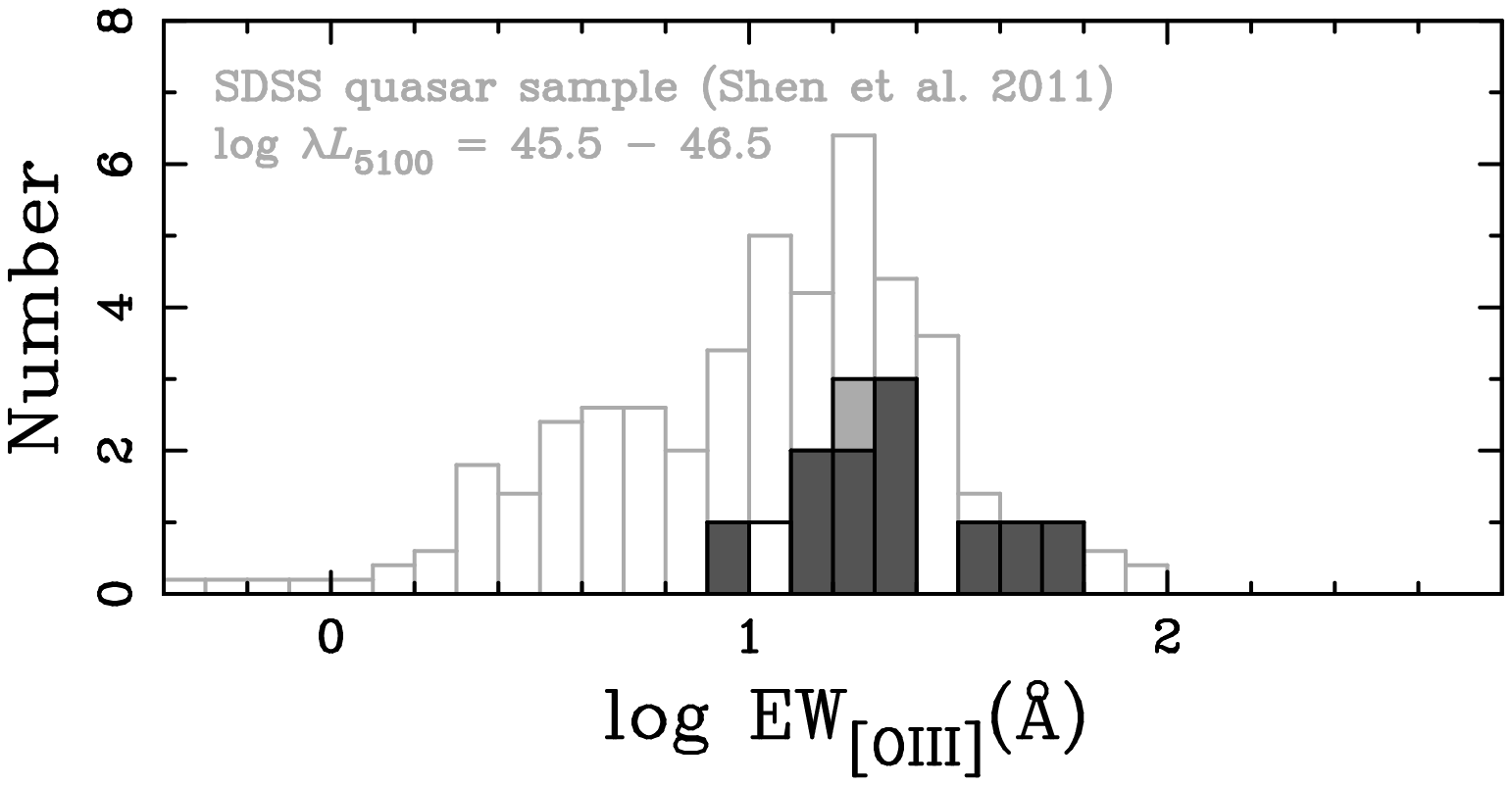}
\caption{Distribution of rest-frame equivalent width of [\ion{O}{iii}]$\lambda$5007 line. Black-line histograms denote our sample of 13 N-loud quasars; dark-grey filled histogram shows our observed 11 N-loud quasars and open histogram shows the 3$\sigma$ upper limit. The result of SDSS J1707$+$6443 \citep{2012A&A...543A.143A} shown as a light-grey filled histogram. Grey-line open histogram denotes the distribution of 232 SDSS quasars at $0 < z < 1$ in a luminosity range at $\log \lambda L_{\rm 5100} =$ 45.5--46.5 \citep{2011ApJS..194...45S}, which is divided by five to compare with our sample easily.}
\label{ewo3}
\end{figure} 

\begin{table} 
\caption{Estimates of 3$\sigma$ flux upper limits of undetected emission lines.}
\label{t:3sig}
\centering
\begin{tabular}{lccc}
\hline\hline
\multicolumn{1}{c}{Object} & \multicolumn{1}{c}{$F_{\rm [\ion{Ne}{v}]}$} & \multicolumn{1}{c}{$F_{\rm [\ion{O}{ii}]}$} & \multicolumn{1}{c}{$F_{\rm [\ion{Ne}{iii}]}$}\\
& \multicolumn{3}{c}{[10$^{-17}$ ergs s$^{-1}$ cm$^{-2}$]}\\
\hline
SDSS J0357$-$0528     & \multicolumn{1}{c}{--$^a$} & $< 16.47$ & $<  9.41$\\
SDSS J0931$+$3439     & \multicolumn{1}{c}{--$^a$} & $< 10.43$ & $<  6.40$\\
SDSS J1211$+$0449     & \multicolumn{1}{c}{--$^a$} & \multicolumn{1}{c}{--$^a$} & \multicolumn{1}{c}{--$^a$}\\
SDSS J0756$+$2054     & $< 17.23$ & $< 12.57$ & $< 14.52$\\
SDSS J0803$+$1742     & $< 17.89$ & $< 13.29$ & $< 12.95$\\
SDSS J0857$+$2636     & $< 11.75$ & $<  7.15$ & $< 10.77$\\
SDSS J0938$+$0900     & $< 28.46$ & $< 18.74$ & $< 24.54$\\
SDSS J1036$+$1247     & $< 21.31$ & $< 15.20$ & $< 14.71$\\
SDSS J1110$+$0456$^b$ & $< 14.39$ & $<  8.80$ & $< 11.52$\\
SDSS J1130$+$1512     & $< 16.26$ & $<  9.87$ & $< 14.31$\\
SDSS J1134$+$1500     & $< 49.02$ & $< 46.02$ & $< 42.98$\\
SDSS J1308$+$1136     & $< 29.67$ & $< 29.45$ & $< 16.50$\\
\hline
\end{tabular}
\tablefoot{$^a$ Out of the wavelength coverage; $^b$ based on the averaged FWHM$_{\rm [\ion{O}{iii}]}$ value of 24.2{\AA}, due to the lack of the [\ion{O}{iii}]$\lambda$5007 line detection.}
\end{table} 

\begin{figure} 
\centering
\includegraphics[width=8.5cm]{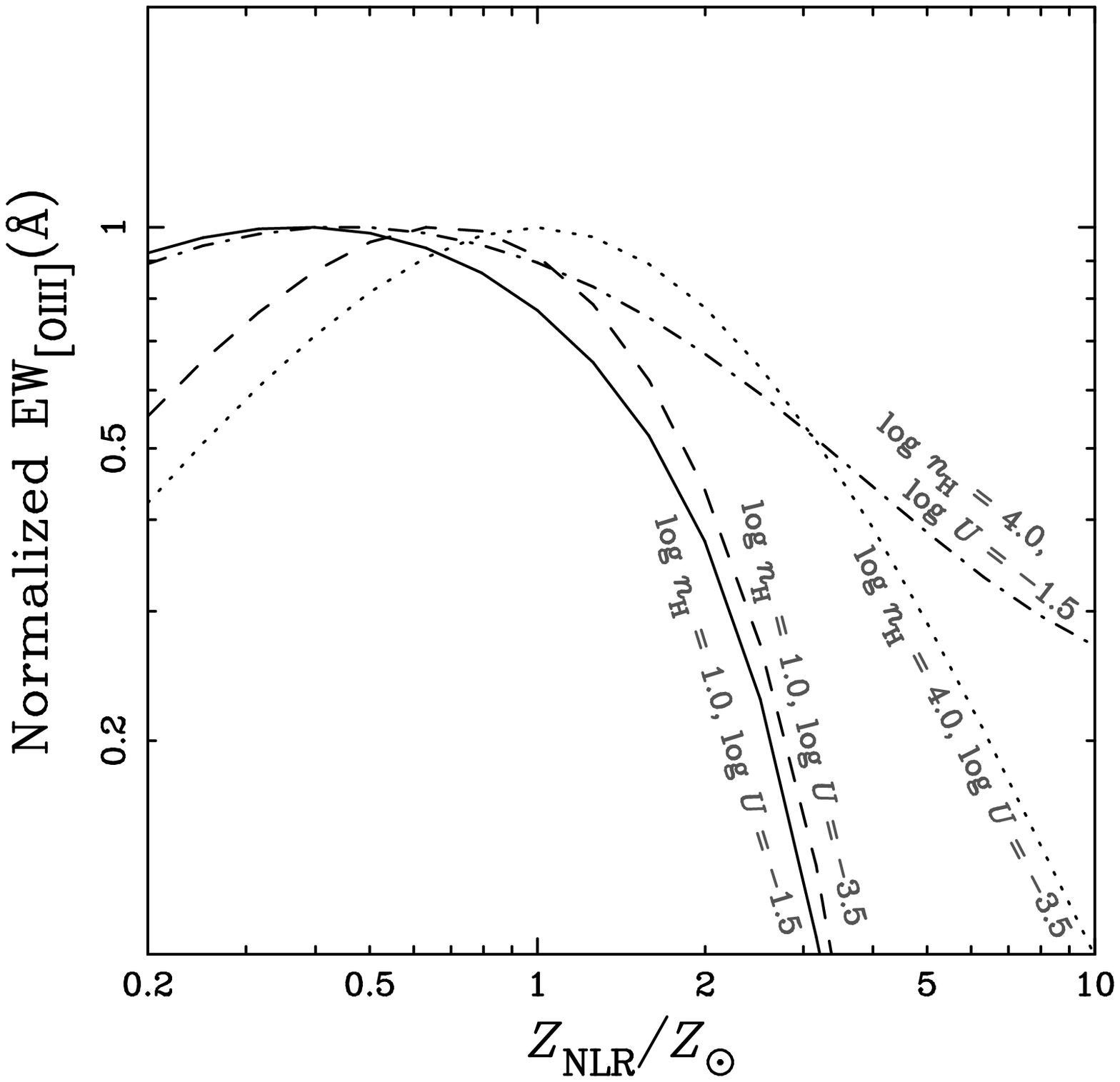}
\caption{Normalized equivalent widths of the [\ion{O}{iii}]$\lambda$5007 line as a function of NLR metallicities, predicted by photoionization models. These are normalized at their respective peaks. Black solid, dashed, dot-dashed, and dot lines are results of photoionization model calculations with ($\log n_{\rm H}$, $\log U$) = (1.0, $-$1.5), (1.0, $-$3.5), (4.0, $-$1.5), and (4.0, $-$3.5), respectively. Given parameters are also labeled for each line.}
\label{cld1}
\end{figure} 

\section{Spectral measurements}\label{spec} 

To measure line properties, we fit emission lines of H$\beta$ and [\ion{O}{iii}]$\lambda$4959,5007 lines with Gaussian functions. In the multi-component fitting processes, we used a routine, {\tt mpcurvefit}, in the Interactive Data Language (IDL). First, we converted our observed spectra to the rest frame. Second, we fit the continuum with a single power-law model. In this fitting, we ignored the host-galaxy starlight and the \ion{Fe}{ii} emission blends due to low signal-to-noise ratios, S/N. The best continuum fit was determined based on the $\chi^2$ statistics in the wavelength ranges where there are no strong emission lines, e.g., 4600 -- 4750{\AA} and 5050 -- 5150\AA\ in the rest frame. From the power-law continuum model, we estimate the monochromatic luminosity at 5100\AA.

After the pseudo-continuum model was subtracted from the original spectrum, we fit emission lines of H$\beta$, [\ion{O}{iii}]$\lambda$4959, and [\ion{O}{iii}]$\lambda$5007 lines. We used one or two Gaussians for the broad and narrow H$\beta$ components. For the narrow [\ion{O}{iii}]$\lambda$4959 and [\ion{O}{iii}]$\lambda$5007 lines, we adopted the single-Gaussian fitting; their flux ratio is fixed as three. For all narrow components, we adopted the same velocity shift and width by assuming the same kinematics of NLR clouds. We measured the emission-line flux of [\ion{O}{iii}]$\lambda$5007 and obtained the rest-frame equivalent width, EW$_{[\ion{O}{iii}]}$, with the continuum flux at 5007{\AA} from the power-law model.

To estimate uncertainties of the line-width and luminosity measurements, we adopted the Monte Carlo flux randomisation method \citep[e.g.,][]{2009ApJ...705..199B,2011ApJS..194...45S,2012ApJ...747...30P}. We generated 1000 mock spectra for each observed spectrum by adding Gaussian random noise based on the flux errors. Then, we measured the line widths and AGN luminosities from the simulated spectra. We derived the standard deviation of their distribution as the measurement uncertainty. We also gave 3$\sigma$ upper limits for [\ion{Ne}{v}]$\lambda$3426, [\ion{O}{ii}]$\lambda$3727, and [\ion{Ne}{iii}]$\lambda$3869 lines by adopting the average velocity width of the [\ion{O}{iii}]$\lambda$5007 emission line ($\Delta v \sim 1450$ km s$^{-1}$), since we could not detect these emission lines in our observations (see Figure~\ref{f:spec}).

In Figure~\ref{f:spec}, some objects (e.g., SDSS J1211$+$0449 and SDSS J1036$+$1247) seem to show blueshifted narrow lines compare to the expected central wavelength indicated by dotted vertical lines \citep{2008ApJ...680..169S}.
Although these velocity shifts may be a sign of gas outflow \citep[e.g.,][]{2005ApJ...632..751R,2014ApJ...795...30B,2016ApJ...817..108W,2017ApJ...837...91B}, it is required to consider the accuracy of redshift measurement carefully to discuss this topic and this is beyond the scope of this paper.
Note that this uncertainty of redshift does not affect our conclusion.

Figure~\ref{f:lfit} shows two examples of the fitting results. Measurement results, i.e., line fluxes and equivalent widths of the [\ion{O}{iii}]$\lambda$5007 line, the full widths at half maximum (FWHM) of the broad H$\beta$ line, and monochromatic luminosities at 5100\AA, are listed in Table~\ref{t:resu}, and 3$\sigma$ upper limits of fluxes for undetected emission lines, i.e., [\ion{Ne}{v}]$\lambda$3426, [\ion{O}{ii}]$\lambda$3727, and [\ion{Ne}{iii}]$\lambda$3869 lines, are shown in Table~\ref{t:3sig}.

\section{Results}\label{resu} 

As described in Section~\ref{obse}, we obtained near-infrared (i.e., $J$ and $H$ bands) spectra of 12 N-loud quasars as shown in Figure~\ref{f:spec}. In $J$-band wavelength, there is no significant detection of emission lines, e.g., [\ion{Ne}{iii}]$\lambda$3869 as \citet{2012A&A...543A.143A} found in the spectra of SDSS J1707$+$6443. On the other hand, $H$-band spectra show significant detections of H$\beta$, [\ion{O}{iii}]$\lambda$4959, and [\ion{O}{iii}]$\lambda$5007 lines for most objects in our sample.

Figure~\ref{ewo3} shows distributions of observed rest-frame equivalent widths of the [\ion{O}{iii}]$\lambda$5007 line of N-loud quasars, denoted as black-line filled histograms. Here, we added SDSS J1707$+$6443 in \citet{2012A&A...543A.143A} to our sample. In order to compare these N-loud quasars with general quasars, we also plot [\ion{O}{iii}]$\lambda$5007 equivalent widths of SDSS quasars at $0 < z < 1$ \citep{2011ApJS..194...45S}, shown as grey-line open histograms in Figure~\ref{ewo3}. Moreover, since some previous studies reported the anti-correlation between the equivalent widths of narrow emission lines and the AGN luminosity \citep[e.g.,][]{1981ApJ...250..469S,2013ApJ...762...51Z}, for consistency in luminosities between N-loud and general quasars, we selected 232 general quasars at $\log \lambda L_{\rm 5100} =$ 45.5 -- 46.5 from the SDSS sample, which is almost the same range as our N-loud quasar sample. As shown in Figure~\ref{ewo3}, we found that the equivalent width of N-loud quasars are slightly larger than that of general SDSS quasars. The averaged equivalent widths (standard deviations) of N-loud and general quasars are 1.34 (0.22) and 0.99 (0.47), respectively. By using the Kolmogorov-Smirnov (KS) test, we checked that there is no significant difference of the EW$_{\rm [\ion{O}{iii}]}$ distributions of them ($p = 0.004$).

In addition to measurements of emission lines, we also obtained black-hole masses and Eddington ratios of our sample shown in Table~\ref{t:resu}, estimated based on the FWHM of the H$\beta$ line and monochromatic luminosity at 5100\AA, which is well calibrated in the local universe \citep[e.g.,][]{2006ApJ...641..689V,2015ApJ...809..123H}. Usually, the single-epoch estimation can be expressed as follows: 
\begin{equation}
\log \left(\frac{M_{\rm BH}}{M_\odot}\right) = a + b \log \left(\frac{\lambda L_\lambda}{10^{44} \ {\rm ergs \ s}^{-1}} \right) + c \log \left(\frac{{\rm FWHM}}{{\rm km \ s}^{-1}} \right).
\end{equation}
In this study, we adopted a recipe of H$\beta$-based black-hole masses in \citet{2006ApJ...641..689V}, i.e., $a = 0.91$, $b = 0.50$, and $c = 2.00$. Eddington ratios are provided through estimates of bolometric luminosities, using a bolometric correction of BC $= 9.26$ \citep{2006ApJS..166..470R,2011ApJS..194...45S}. Note that we gave only statistical errors of black-hole masses and Eddington ratios in Table~\ref{t:resu}, excluding systematic ones originated from calibrations and virial parameters. In addition to the near-infrared spectra, our objects have observed optical spectra obtained with SDSS spectrographs \citep{2008ApJ...680..169S,2011ApJS..194...45S}. Thus, we collected the rest-frame UV measurements, i.e., FWHM$_\ion{C}{iv}$, $\log (\lambda L_{\rm 1350})$, $\log (M_{\rm BH,\ion{C}{iv}})$, and $\log (L/L_{\rm Edd})_{\rm \ion{C}{iv}}$, in Table~\ref{t:sdss}. Here, \ion{C}{iv}$\lambda$1549-based black-hole masses is estimated with a recipe of $a = 0.66$, $b = 0.53$, and $c = 2.00$ \citep{2006ApJ...641..689V}. In the estimation of bolometric luminosities from the monochromatic luminosity at 1350\AA, we adopted BC $= 3.81$ \citep{2011ApJS..194...45S}. For the 12 N-loud quasars in our sample, measurements of FWHMs of the \ion{C}{iv}$\lambda$1549 line and monochromatic luminosities at 1350{\AA} are available from \citet{2011ApJS..194...45S}.

Note that we ignore the effect of quasar variability on the results for black hole masses and Eddington ratios, since a typical luminosity variability amplitude of quasars is only $\sim 0.2$ magnitude on timescales up to several years \citep[e.g.,][]{2007AJ....134.2236S,2010ApJ...721.1014M,2012ApJ...753..106M} and the uncertainty of the single-epoch masses would be dominated by measurement errors (see \citealt{2013BASI...41...61S} for more detail).

\section{Discussion}\label{disc} 

In this section, we discuss chemical and physical parameters of N-loud quasars by using the measurements as described in Sections~\ref{spec} and \ref{resu}. First, we investigate their NLR metallicities with a diagnostic method consisting of equivalent width of [\ion{O}{iii}]$\lambda$5007 line \citep{2012A&A...543A.143A}. Second, we discuss the virial black-hole mass of N-loud quasars estimated based on the H$\beta$ line (see Sec.~\ref{resu}). We compare these black-hole masses with those based on rest-frame ultraviolet spectra, i.e., the broad \ion{C}{iv}$\lambda$1549 line. We also show their Eddington ratios, and discuss a possibility about the connection between AGN and star formation.

\subsection{NLR metallicities} 

The emission line of [\ion{O}{iii}]$\lambda$5007 is a collisionally-excited line and the emissivity depends strongly on the gas temperature. The equilibrium temperature of ionized gas clouds is significantly affected by gas metallicity, because metal emission lines are main coolants of these clouds. Therefore, collisionally-excited lines are good indicators of NLR metallicities; if the NLR metallicity is high, these emission lines should be faint.

To see the metallicity dependence of the [\ion{O}{iii}]$\lambda$5007 equivalent width for NLR clouds, we used the photoionization code Cloudy \citep[version 13.05;][]{1998PASP..110..761F,2013RMxAA..49..137F}. In this calculation, we adopted basically the same parameters as those in \citet{2012A&A...543A.143A}, i.e., the hydrogen densities of $n_{\rm H} = 10^{1.0}$ and $10^{4.0}$ cm$^{-3}$, the ionization parameters of $U = 10^{-3.5}$ and $10^{-1.5}$, which parameter ranges are typical of low-$z$ AGNs: \citet{2012MNRAS.427.1266V} determined densities and ionization parameters of local Seyfert galaxies based on optical emission-line diagnostic diagrams \citep[see also][]{2006A&A...456..953B,2006A&A...459...55B,2001ApJ...546..744N,2002ApJ...575..721N}. For the spectral energy distribution (SED) of the photionizing continuum radiation \citep{1987ApJ...323..456M}, we adopted the "table AGN" command, which is reproduces the typical SED of AGNs. Regarding the chemical composition of NLR gas, we assumed dusty gas clouds contained Orion-type graphite and silicate grains, that all metals are changed basically with a solar abundance ratio but we also consider the metal depletion onto dust grains (see Table~7.7 in Hazy, a brief introduction to Cloudy C13.05) except for helium and nitrogen. We used analytical expressions for He and N relative abundances as functions of metallicities given in \citet{2000ApJ...542..224D}: for helium, we consider both a primary nucleosynthesis component and the primordial value \citep{1992ApJ...384..508R} and nitrogen is assumed to be a secondary nucleosynthesis element \citep[e.g.,][]{1998ApJ...497L...1V}. The metal depletion is also considered to these elements. We stoped the calculations when the hydrogen ionization fraction drops below 15\%.

Figure~\ref{cld1} shows the simulation result of EW$_{\rm [\ion{O}{iii}]}$ as a function of the NLR metallicity. This confirms that equivalent width of the [\ion{O}{iii}]$\lambda$5007 line decreases rapidly as the metallicity increases, which is quite small at $Z_{\rm NLR} > 5 Z_\odot$. These results are consistent with predictions obtained by using the Cloudy version 08.00 in \citet{2012A&A...543A.143A}. With this method, we can examine whether NLR metallicities of N-loud quasars are extremely high (i.e., $Z_{\rm NLR} > 5 Z_\odot$) relative to general quasars, or not, although we can not estimate accurate metallicities only by using the [\ion{O}{iii}]$\lambda$5007 equivalent width. As described in Section~\ref{resu}, we found that there is no significant difference between N-loud quasars and typical SDSS quasars in the same luminosity range. Note that since NLR metallicities show no significant redshift evolution \citep{2006A&A...447..157N,2009A&A...503..721M}\footnote{It has been sometimes reported that the AGN metallicity shows no redshift evolution, that is inferred for BLRs \citep[e.g.,][]{2006A&A...447..157N}, not only for NLRs. Note that significant redshift evolution of the gas metallicity is reported for star-forming galaxies \citep[e.g.,][]{2008A&A...488..463M}, damped Ly$\alpha$ \citep[e.g.,][]{2012ApJ...755...89R}, and sub-damped Ly$\alpha$ \citep[e.g.,][]{2003MNRAS.345..480P,2015ApJ...806...25S}. The origin of the discrepancy in the metallicity evolution between galaxies and AGNs (or the lack of the significant metallicity evolution in AGNs) has been discussed in, e.g., \citet{2009A&A...494L..25J}.}, we are able to compare our sample at $z \sim 2.2$ with the SDSS quasars at $0 < z < 1$. This result, therefore, indicates that NLR metallicities of N-loud quasars are not extremely high relative to normal ones. This result confirms the result obtained by SDSS J1707$+$6443 in \citet{2012A&A...543A.143A}.

\begin{figure*} 
\centering
\includegraphics[width=5.9cm]{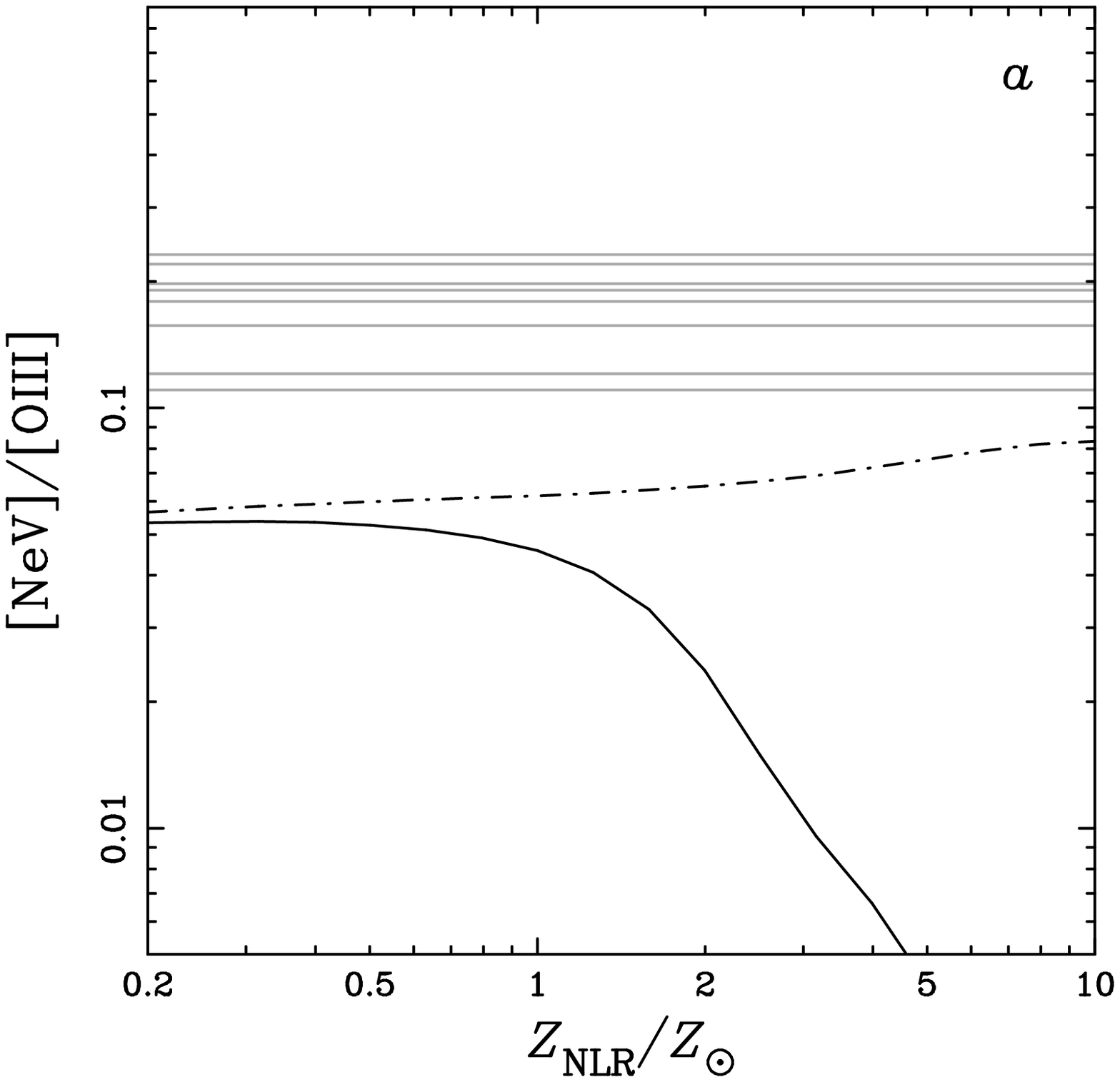}
\hspace{0.1cm}
\includegraphics[width=5.9cm]{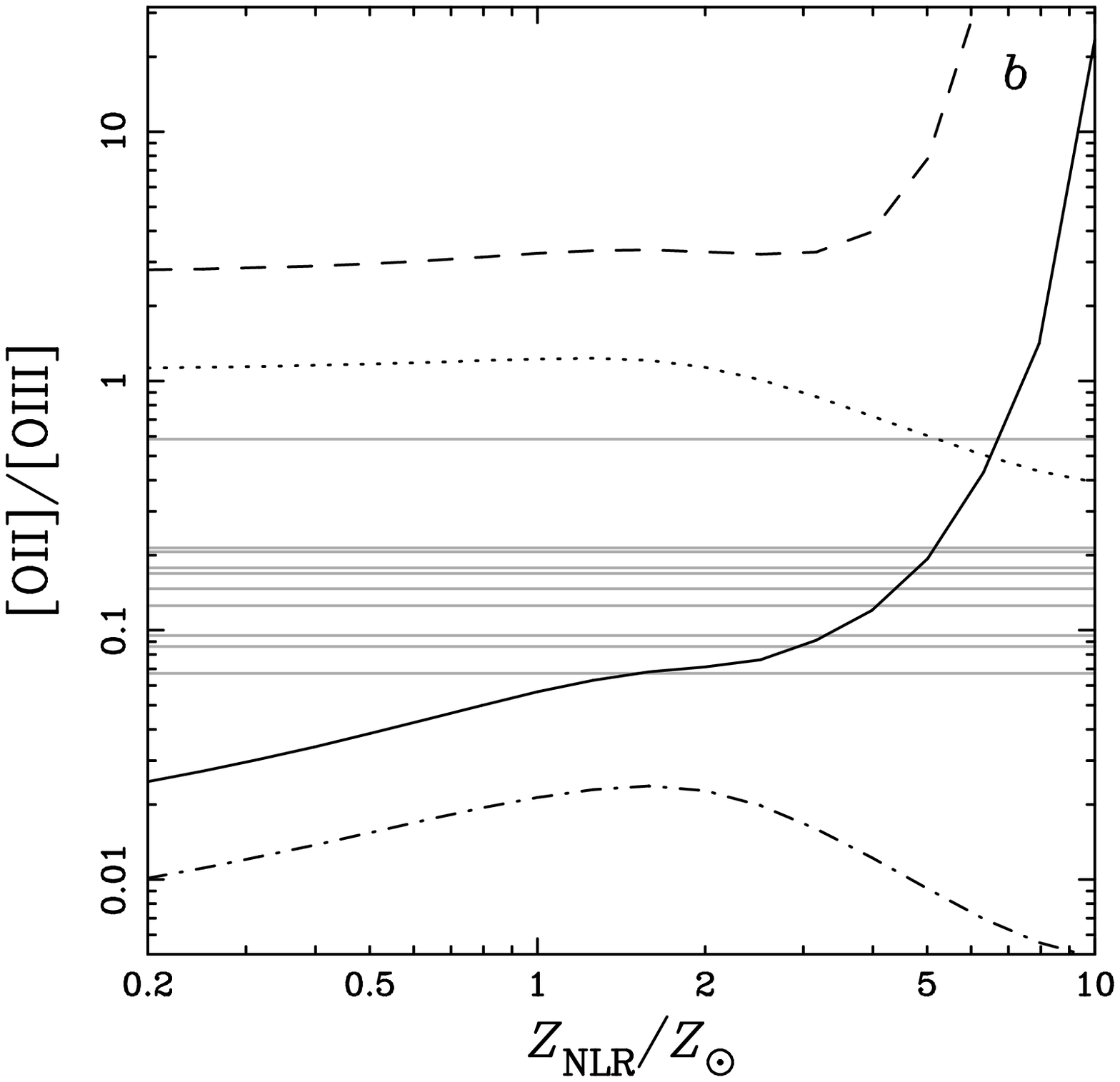}
\hspace{0.1cm}
\includegraphics[width=5.9cm]{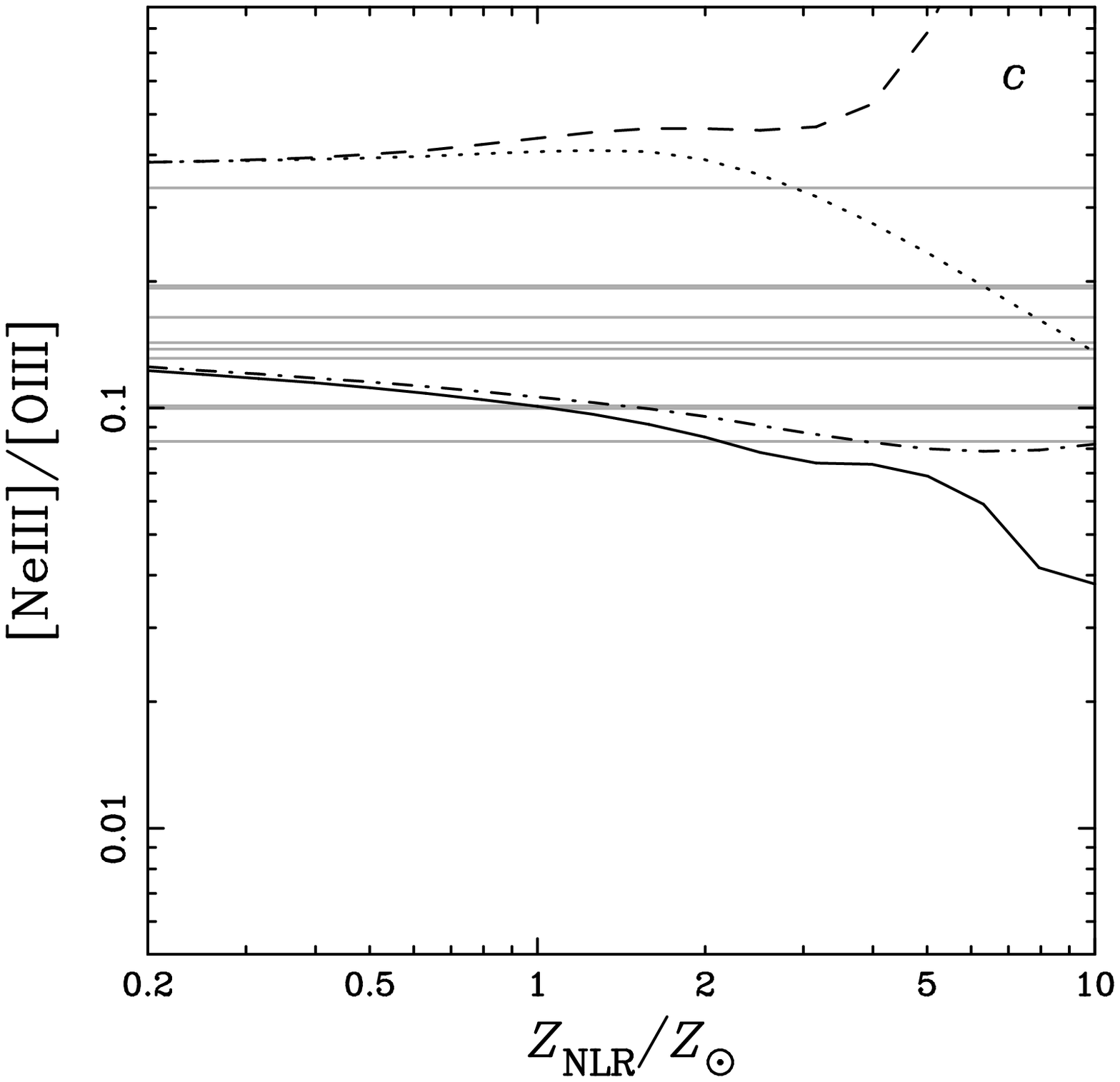}
\caption{Model predictions of emission-line flux ratios, i.e., [\ion{Ne}{v}]$\lambda$3426/[\ion{O}{iii}]$\lambda$5007 (left, $a$), [\ion{O}{ii}]$\lambda$3727/[\ion{O}{iii}]$\lambda$5007 (middle, $b$), and [\ion{Ne}{iii}]$\lambda$3869/[\ion{O}{iii}]$\lambda$5007 (right, $c$), as a function of NLR metallicities. Black solid, dashed, dot-dashed, and dot lines are results of photoionization model calculations with ($\log n_{\rm H}$, $\log U$) = (1.0, $-$1.5), (1.0, $-$3.5), (4.0, $-$1.5), and (4.0, $-$3.5), respectively. Note that the left panel shows only two high-$U$ models because [\ion{Ne}{v}]$\lambda$3426 line fluxes in low-$U$ models are too faint relative to [\ion{O}{iii}]$\lambda$5007 fluxes. Horizontal lines show the 3$\sigma$ upper limits of our sample.}
\label{cldy}
\end{figure*} 

As described in Section~\ref{resu}, we also measured 3$\sigma$ flux upper limits of undetected emission lines, i.e., [\ion{Ne}{v}]$\lambda$3426, [\ion{O}{ii}]$\lambda$3727, and [\ion{Ne}{iii}]$\lambda$3869, listed in Table~\ref{t:3sig}. Thus, to check whether the result obtained by the [\ion{O}{iii}]$\lambda$5007 equivalent width is consistent with these upper limits, we compare the observational upper limits with model predictions obtained with the model calculations. Figure~\ref{cldy} shows the emission line flux ratios, i.e., [\ion{Ne}{v}]$\lambda$3426/[\ion{O}{iii}]$\lambda$5007, [\ion{O}{ii}]$\lambda$3727/[\ion{O}{iii}]$\lambda$5007, and [\ion{Ne}{iii}]$\lambda$3869/[\ion{O}{iii}]$\lambda$5007, as a function of NLR metallicities predicted by Cloudy. As shown in Figure~\ref{cldy}$a$, we found that the predicted emission line flux ratios of [\ion{Ne}{v}]$\lambda$3426/[\ion{O}{iii}]$\lambda$5007 are below the observational upper limits (grey-horizontal lines) for all of the parameter sets examined in our calculations. Note that the predicted [\ion{Ne}{v}]$\lambda$3426/[\ion{O}{iii}]$\lambda$5007 flux ratio of the models with $\log U = -3.5$ is far below the range shown in Figure~\ref{cldy}$a$. Figure~\ref{cldy}$b$ shows the line flux ratio of [\ion{O}{ii}]$\lambda$3727/[\ion{O}{iii}]$\lambda$5007 as a function of NLR metallicity; this line flux ratio is sensitive to ionization parameters. This figure implies that it is difficult to explain the observational results with low-$U$ models, i.e., $\log U = -3.5$. On the other hand, models with high ionization parameters satisfy most of observed upper limits at $Z_{\rm NLR} < 4 Z_\odot$, whereas at $Z_{\rm NLR} > 5 Z_\odot$, the low-$n_{\rm H}$ and high-$U$ model (a solid line) is restricted by all observational upper limits. This is consistent with the result of the [\ion{O}{iii}]$\lambda$5007 equivalent width, that NLR metallicities of N-loud quasars are not extremely high. Moreover, in Figure~\ref{cldy}$c$ we also confirmed that most objects can be explained with high-$U$ model predictions of [\ion{Ne}{iii}]$\lambda$3869/[\ion{O}{iii}]$\lambda$5007 ratio instead of low-$U$ models. Note that it seems to be difficult to discuss hydrogen densities with the upper limits. As a result, we confirmed that these observational upper limits are consistent with the scenario that NLR metallicities of N-loud quasars are not extremely high.

In this section, we found that NLR clouds of N-loud quasars are not extremely metal rich generally. To interpret this observational result, we will see two possible scenarios as follows. One is that the metallicity of N-loud quasars are not very high in both BLRs and NLRs, assuming the correlation between $Z_{\rm BLR}$ and $Z_{\rm NLR}$. In this case, our result indicates that BLR clouds of N-loud quasars are characterized by a very high nitrogen relative abundance without extremely high $Z_{\rm BLR}$. This means broad-emission lines of nitrogen are unreliable indicators of the BLR metallicity in these cases. Another possibility is that $Z_{\rm BLR}$ is determined by local enrichment at the BLR scale, without an effect of global metallicity in their host galaxies traced by NLR metallicities, because the total mass of BLR clouds is typically very low \citep[$M_{\rm BLR} \sim 10^2$ -- $10^4$;][]{2003ApJ...582..590B}. This picture can explain the situation that $Z_{\rm NLR}$ is not very high while $Z_{\rm BLR}$ is extremely high in N-loud quasars. In this case, since BLR metallicities do not trace chemical properties in galactic scale, we should use metallicities of NLR gases instead of BLR metallicities if we want to investigate chemical status in their host galaxies, especially at high redshift. Note, however, that \citet{2008ApJ...679..962J} shows that emission line flux ratios excluding nitrogen lines of N-loud quasars seem not to be different from those of normal quasars. This means that BLR metallicities do not have metal rich gas clouds. Therefore, we conclude that BLR and NLR metallicities of N-loud quasars are not extremely high, and nitrogen relative abundance is individually high as the former scenario. We will discuss a reason why N-loud quasars show BLR gas clouds containing high N relative abundance by considering a connection between BH growth and nitrogen enrichment via a star formation in the following sections. 

\begin{figure*} 
\centering
\includegraphics[width=5.9cm]{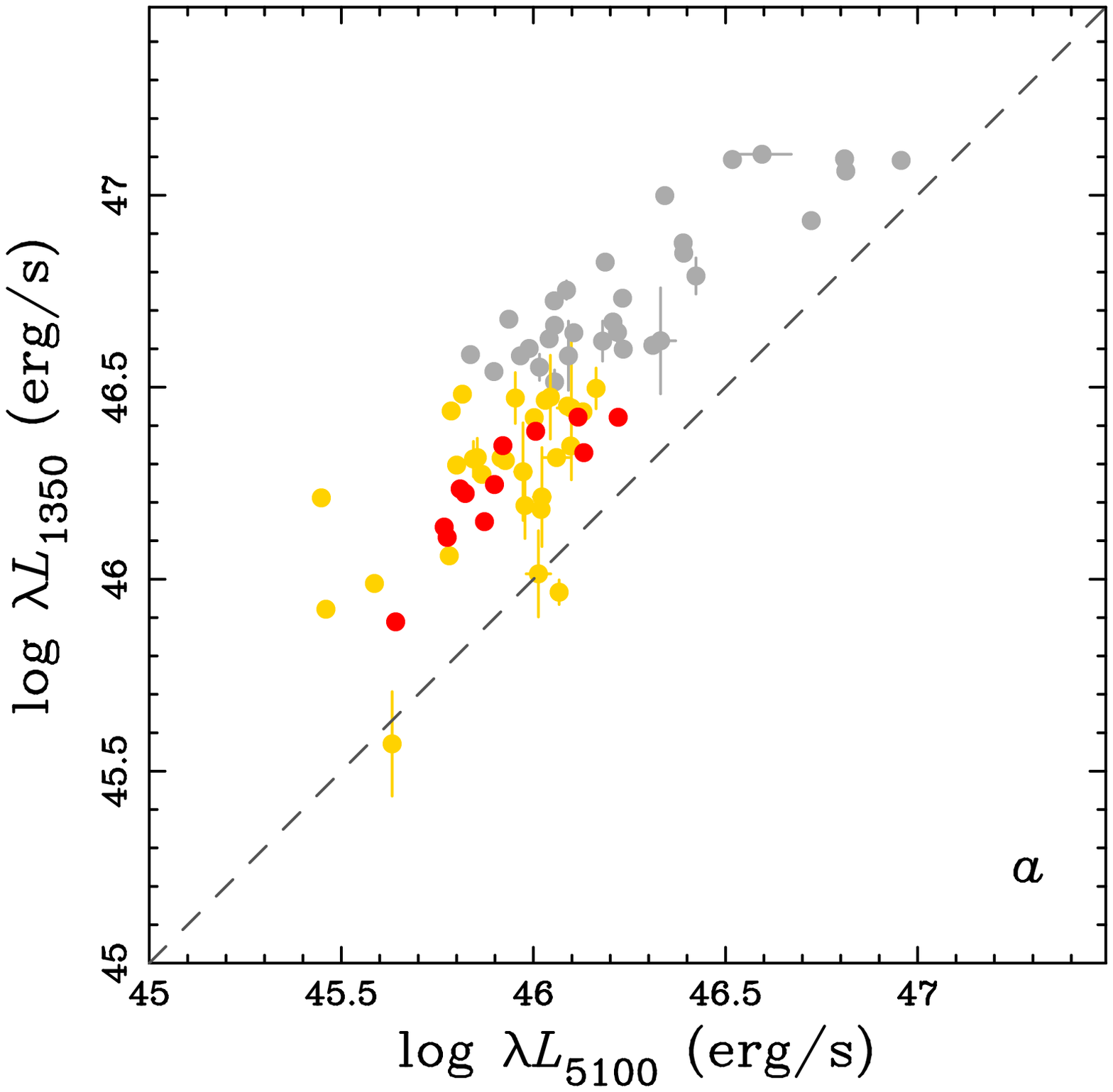}
\hspace{0.1cm}
\includegraphics[width=5.9cm]{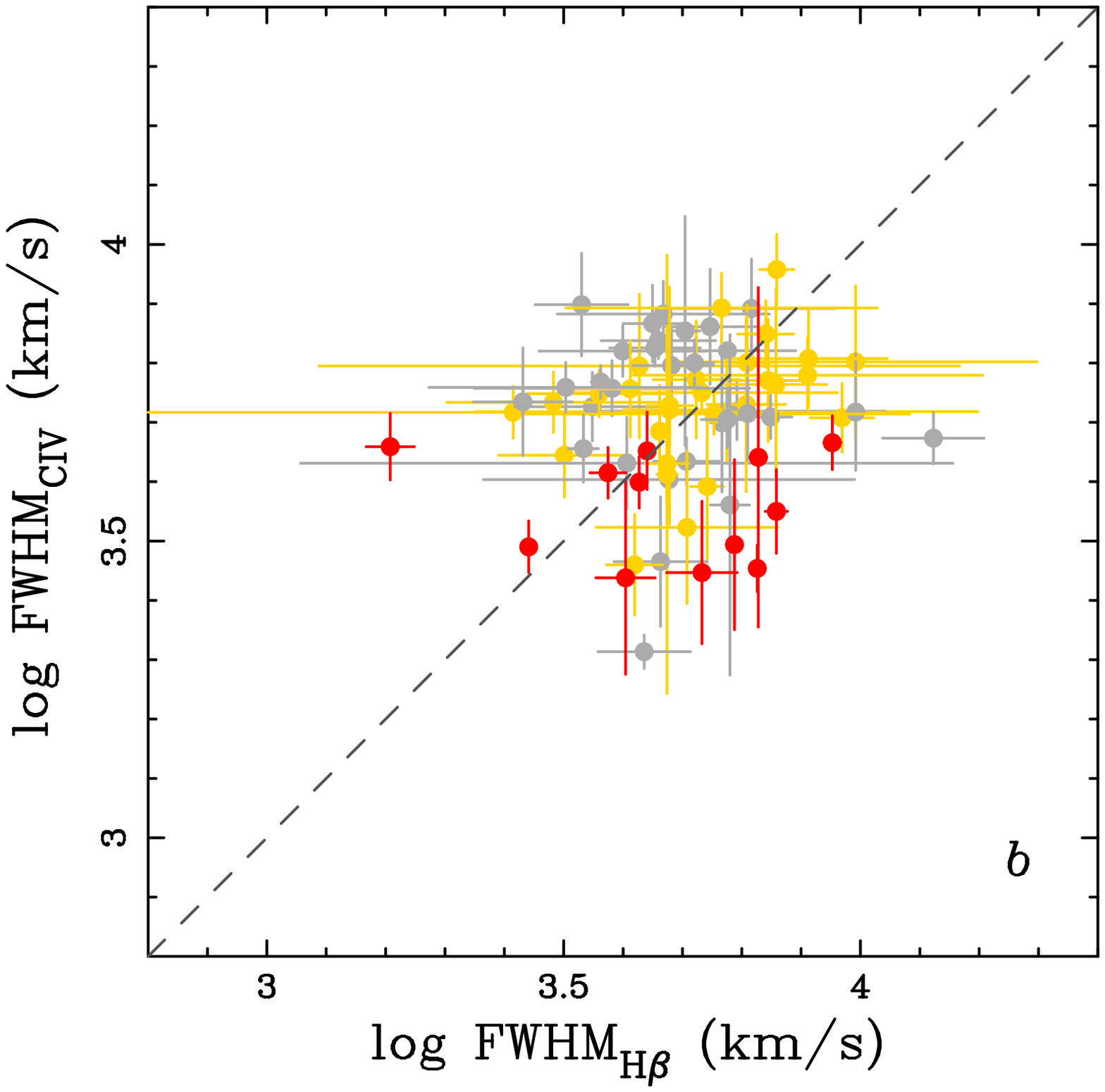}
\hspace{0.1cm}
\includegraphics[width=5.9cm]{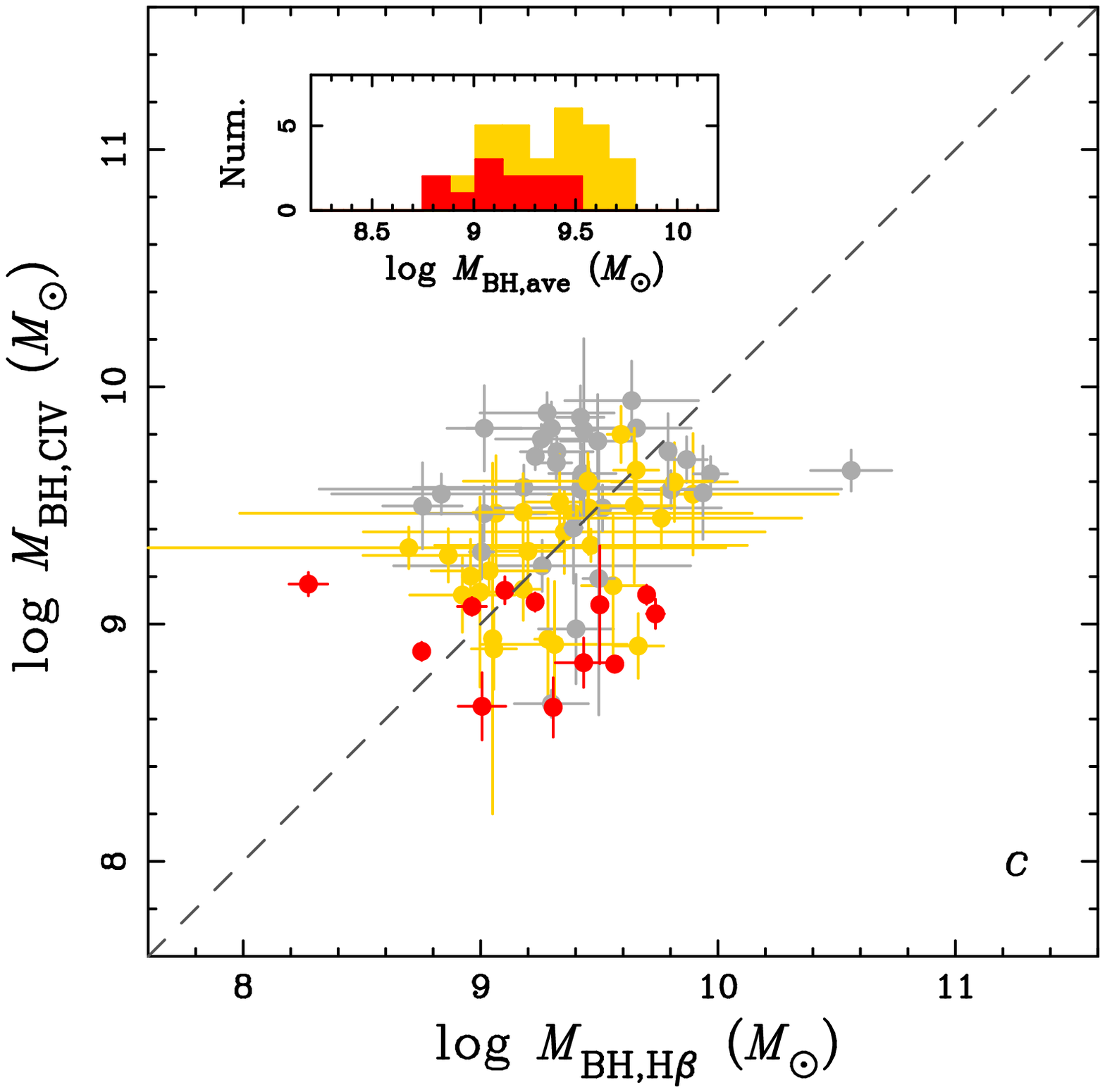}
\caption{Comparison of the monochromatic luminosities at 1350{\AA} and 5100{\AA} (left, $a$), the FWHM of the H$\beta$ line and the \ion{C}{iv}$\lambda$1549 line (middle, $b$), and virial black-hole masses based on \ion{C}{iv}$\lambda$1549 line and H$\beta$ line (right, $c$). Our sample of N-loud quasars are shown as red circles. Grey and yellow circles denote SDSS normal quasars at $1.5 < z < 2.2$ in luminosity ranges of $\log \lambda L_{1350} > 46.5$ and $\log \lambda L_{1350} < 46.5$ from \citet{2012ApJ...753..125S}. Errors represent 1$\sigma$ uncertainties of measurements, excluding systematic errors. The one-to-one relations are indicated by dashed lines for each panel.}
\label{mass}
\end{figure*} 

\subsection{Black-hole masses} 

As described in Section~\ref{resu}, we estimated black-hole masses of N-loud quasars based on the rest-frame optical measurements, i.e., FWHM of H$\beta$ line and continuum luminosities at 5100\AA, and the rest-frame UV measurements, i.e., FWHM$_{\rm \ion{C}{iv}}$ and $L_{\rm 1350}$. In this subsection, we first examine the relation between $L_{\rm 5100}$ and $L_{\rm 1350}$, and the FWHM$_{\rm H\beta}$ and FWHM$_{\rm \ion{C}{iv}}$ relation. Then we compare the black hole masses of $M_{\rm BH,H\beta}$ and $M_{\rm BH,\ion{C}{iv}}$.

Figure~\ref{mass}$a$ shows a relation of monochromatic luminosities at 5100{\AA} and 1350{\AA} of N-loud quasars, shown as red circles. For comparison, we also plotted SDSS general quasars at $1.5 < z < 2.2$ in \citet{2012ApJ...753..125S}, represented with grey and yellow circles. In the result, we confirmed that N-loud quasars follow the $L_{5100\AA}$ and $L_{1350\AA}$ relation of general quasars. As shown in this figure, our N-loud quasars are distributed below the 1350\AA\ luminosity of $\log \lambda L_{\rm 1350} \sim 46.5$. Thus, to compare them to our N-loud quasars without possible luminosity effects, hereafter we colored the quasars at $\log \lambda L_{\rm 1350} > 46.5$ and $\log \lambda L_{\rm 1350} < 46.5$ with grey and yellow, respectively. Figure~\ref{mass}$b$ shows the comparison of FWHMs of H$\beta$ and \ion{C}{iv}$\lambda$1549 lines. It is apparently shown that FWHM$_\ion{C}{iv}$ of N-loud quasars is narrower than those of general quasars typically, although FWHM$_{\rm H\beta}$ seems not to show significant difference between N-loud and normal quasars: the mean values (standard deviations) of FWMH$_\ion{C}{iv}$ and FWHM$_{\rm H\beta}$ are 3692 km s$^{-1}$ (715 km s$^{-1}$) and 5159 km s$^{-1}$ (1992 km s$^{-1}$) for N-loud quasars, whereas those are 5507 km s$^{-1}$ (1229 km s$^{-1}$) and 5652 km s$^{-1}$ (1770 km s$^{-1}$) for general quasars. This is consistent with the claim that the line widths of carbon broad emission lines of N-loud quasars are narrower than those of general quasars \citep{2008ApJ...679..962J,2004AJ....128..561B}. Many studies claimed that high-ionized BLRs with \ion{C}{iv}$\lambda$1549 line are affected by a non-virial component such as outflow, and would be a biased mass estimator \citep[e.g.,][]{2012ApJ...753..125S,2012NewAR..56...49M}. Note that, however, H$\beta$ line widths obtained in this study have also uncertainties due to low S/N of near-infrared spectroscopies, implying that black-hole masses estimated both with the H$\beta$ and \ion{C}{iv}$\lambda$1549 lines should involve respective uncertainties. Hence, hereafter we regard mean black-hole masses between them as their representative black-hole masses. As shown in Figure~\ref{mass}$c$, we found that black-hole masses of N-loud quasars seem to be lower than those of general quasars even by focusing on the same luminosity objects shown as yellow circles: the averaged masses are 9.218 and 9.408 logarithmically for N-loud and general quasars, respectively: averaged masses estimated by H$\beta$ line are 9.214 and 9.327, and the averaged masses estimated by \ion{C}{iv}$\lambda$1549 line are 8.965 and 9.320, respectively. This implies that N-loud quasars are situated in a BH growth phase although we should keep their large uncertainties of BH masses in mind, and thus we discuss mass accretions into SMBHs with nitrogen enrichment of N-loud quasars.

\begin{figure} 
\centering
\includegraphics[width=8.5cm]{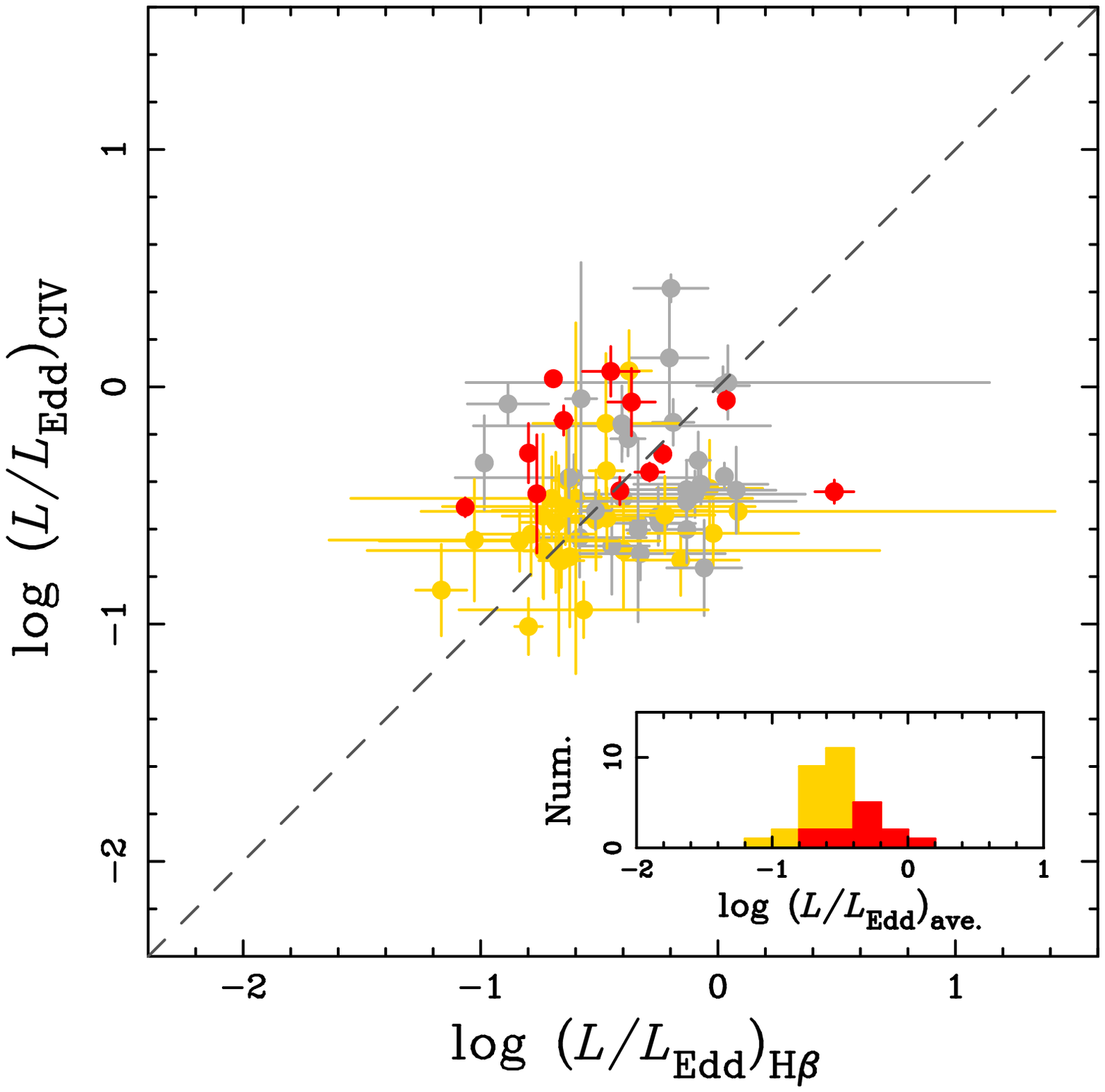}
\caption{Comparison of Eddington ratios estimated by the H$\beta$ line with that by the \ion{C}{iv}$\lambda$1549 line. Red, grey, and yellow circles are the same as those in Figure~\ref{mass}. The one-to-one relation is indicated by the dashed line. Eddington ratios averaged from H$\beta$ and \ion{C}{iv}$\lambda$1549 lines are given as histograms on the small bottom panel. The red and yellow-filled histograms show averaged Eddington ratios of our N-loud quasars and SDSS normal ones at $\log \lambda L_{1350} < 46.5$ from \citet{2012ApJ...753..125S}, respectively.}
\label{eddi}
\end{figure} 

\subsection{AGN activities of N-loud quasars} 

In Section~5.1, we found that NLR clouds in N-loud quasars are not extremely metal rich, indicating a picture which only nitrogen relative abundance of the BLR gas clouds is relatively high in the N-loud quasars \citep[e.g.,][]{2008ApJ...679..962J,2012A&A...543A.143A}. In this section, we discuss an interpretation of this nitrogen excess by considering the connection between AGN activity and the nitrogen enrichment due to the nuclear star formation.

Previously, by using optical spectra of 2383 SDSS quasars at $2.3 < z < 3.0$, we investigated the dependence of the BLR metallicity on the Eddington ratio \citep{2011A&A...527A.100M}, where we utilized emission-line flux ratios of \ion{N}{v}$\lambda$1240/\ion{C}{iv}$\lambda$1549, \ion{N}{v}$\lambda$1240/\ion{He}{ii}$\lambda$1640, (\ion{Si}{iv}$\lambda$1398+\ion{O}{iv}]$\lambda$1402)/\ion{C}{iv}$\lambda$1549, and \ion{Al}{iii}$\lambda$1857/\ion{C}{iv}$\lambda$1549, that are sensitive to the BLR metallicity. In the result, we found that the Eddington ratio depends on the emission-line flux ratios involving the \ion{N}{v}$\lambda$1240 line, while it does not correlate with the other emission-line flux ratios, i.e., (\ion{Si}{iv}$\lambda$1398+\ion{O}{iv}]$\lambda$1402)/\ion{C}{iv}$\lambda$1549 and \ion{Al}{iii}$\lambda$1857/\ion{C}{iv}$\lambda$1549 (see Figure~6 in \citealt{2011A&A...527A.100M} for more detail). We suggested that the correlation between the Eddington ratio and the line ratios including nitrogen, is tracing a delay of the mass accretion into SMBHs relative to the onset of nuclear star formation of about 10$^8$ years reported by \citet{2007ApJ...671.1388D}, which is the time scale required for the nitrogen enrichment from AGB stars. Thus, in other words, this result indicates that the mass loss from AGB stars enrich BLR gas clouds with nitrogen and enhance the accretion into SMBHs.

\begin{table*} 
\caption{Measurements of rest-frame UV spectra obtained by SDSS observations.}
\label{t:sdss}
\centering
\begin{tabular}{lcccr}
\hline\hline
\multicolumn{1}{c}{Object} & $\log$FWHM$_{\rm \ion{C}{iv}}$ & $\log (\lambda L_{\rm 1350})$ & $\log M_{\rm BH,\ion{C}{iv}}$ & \multicolumn{1}{c}{$\log (L/L_{\rm Edd})_{\rm \ion{C}{iv}}$}\\
& [km s$^{-1}$] & \multicolumn{1}{c}{[ergs s$^{-1}$]} & [$M_\odot$] &\\
\hline
SDSS J0357$-$0528 & 3.666$\pm$0.045 & 46.136$\pm$0.011 & 9.124$\pm$0.040 & $-$0.507$\pm$0.040\\
SDSS J0931$+$3439 & 3.494$\pm$0.143 & 45.889$\pm$0.019 & 8.649$\pm$0.125 & $-$0.280$\pm$0.125\\
SDSS J1211$+$0449 & 3.641$\pm$0.286 & 46.150$\pm$0.013 & 9.082$\pm$0.249 & $-$0.451$\pm$0.249\\
SDSS J0756$+$2054 & 3.438$\pm$0.163 & 46.109$\pm$0.017 & 8.654$\pm$0.142 & $-$0.064$\pm$0.142\\
SDSS J0803$+$1742 & 3.454$\pm$0.039 & 46.385$\pm$0.008 & 8.832$\pm$0.034 &    0.034$\pm$0.034\\
SDSS J0857$+$2636 & 3.447$\pm$0.120 & 46.422$\pm$0.007 & 8.838$\pm$0.104 &    0.065$\pm$0.104\\
SDSS J0938$+$0900 & 3.599$\pm$0.044 & 46.330$\pm$0.010 & 9.093$\pm$0.038 & $-$0.283$\pm$0.021\\
SDSS J1036$+$1247 & 3.659$\pm$0.056 & 46.246$\pm$0.009 & 9.169$\pm$0.049 & $-$0.442$\pm$0.049\\
SDSS J1110$+$0456 & 3.652$\pm$0.065 & 46.223$\pm$0.015 & 9.143$\pm$0.057 & $-$0.439$\pm$0.057\\
SDSS J1130$+$1512 & 3.490$\pm$0.044 & 46.348$\pm$0.007 & 8.885$\pm$0.038 & $-$0.057$\pm$0.038\\
SDSS J1134$+$1500 & 3.615$\pm$0.043 & 46.235$\pm$0.014 & 9.075$\pm$0.038 & $-$0.359$\pm$0.038\\
SDSS J1308$+$1136 & 3.550$\pm$0.071 & 46.422$\pm$0.007 & 9.044$\pm$0.062 & $-$0.141$\pm$0.045\\
\hline
\end{tabular}
\end{table*} 

By considering the relation between the Eddington ratio and nitrogen abundance described above, we lead one possible scenario that N-loud quasars are situated in the AGN active phase, i.e., high accretion phase. If this scenario is correct, N-loud quasars should show relatively high Eddington ratios. Thus, we calculated Eddington ratios of N-loud quasars and plotted them in Figure~\ref{eddi}. For comparison, we also plotted general quasar sample of \citet{2012ApJ...753..125S} same as Figure~\ref{mass}. To examine the distribution of Eddington ratios of N-loud and general quasars, we calculated averaged Eddington ratios from H$\beta$-based and \ion{C}{iv}$\lambda$1549-based black-hole masses. The distributions of averaged Eddington ratios of our N-loud sample are shown in Figure~\ref{eddi} as red-filled histograms. Regarding the general quasars at $\log \lambda L_{\rm 1350} < 46.5$, we derived averaged Eddington ratios with the same way, shown as the yellow histogram. The mean values (standard deviations) of their Eddington ratios are $-$0.28 (0.24) and $-$0.54 (0.22) for our N-loud quasars and general ones, respectively. It indicates that Eddington ratios of N-loud quasars seems to be slightly higher than those of SDSS quasars: the KS test confirms the significant difference between their distributions with $p = 0.006$. This result means N-loud quasars are situated in a rapid accretion phase significantly, consistent with the scenario that AGB stars enrich BLRs with nitrogen and enhance the accretion into SMBHs.

\section{Conclusions}\label{conc} 

We have performed near-infrared spectroscopic observations of 12 N-loud quasars at $2.09 < z < 2.57$ to investigate their rest-frame optical properties. In the result, we found the following results.
\begin{enumerate}
\item
The rest-frame optical spectra show that [\ion{O}{iii}]$\lambda$5007 equivalent widths of N-loud quasars are slightly larger than those of general quasars, suggesting that NLR metallicities of N-loud quasars are not extremely high as expected with strong broad nitrogen lines. If BLR metallicities are related to chemical properties in NLRs, this result indicates a scenario that nitrogen broad lines depend not only on the BLR metallicity but also other parameters in some cases.
\item
We found N-loud quasars seem to show higher Eddington ratios than those of general quasars, indicating that the N-loud quasars are in a high accretion phase. This result can be predicted from an observational positive correlation between Eddington ratios and nitrogen abundance found in \citet{2011A&A...527A.100M}, suggesting a possible connection between AGN activities and star formation where the mass accretion onto SMBH is associated with a post-starburst phase, when AGB stars enrich the interstellar medium with nitrogen.
\end{enumerate}
To understand not only the chemical properties of high-$z$ galaxies but also the connection between SMBH growths and galaxy formations, investigating chemical compositions of BLR gas is crucial issues. Especially, it is important to examine whether nitrogen broad lines are really good indicators of BLR metallicities or not, because they are often used as BLR metallicity indicators. In this sense, understanding backgrounds of N-loud quasars would be helpful as a first step, since N-loud quasars would show us an extreme case regarding chemical evolution and black-hole growth.

\begin{acknowledgements}
We would like to thank the anonymous referee for useful comments and suggestions.
We are also grateful to Nobuo Araki and Yosuke Minowa for observational supports, and thank NTT and Subaru telescope staffs for their invaluable help during the observations.
We would like to thank Gary J. Ferland for providing the great photoionization code Cloudy.
K.M. acknowledges financial support from the Japan Society for the Promotion of Science (JSPS) KAKENHI Grant No. 14J01811.
T.N. is financially supported by the JGC-S Scholarship Foundation and JSPS KAKENHI Grant No. 25707010, 16H01101, and 16H03958.
R.M. acknowledges ERC Advanced Grant 695671 ``QUENCH'' and support by the Science and Technology Facilities Council (STFC).
D.P. acknowledges support through the EACOA Fellowship from The East Asian Core Observatories Association, which consists of the National Astronomical Observatories, Chinese Academy of Science (NAOC), the National Astronomical Observatory of Japan (NAOJ), Korean Astronomy and Space Science Institute (KASI), and Academia Sinica Institute of Astronomy and Astrophysics (ASIAA).
Data analysis were in part carried out on common-use data analysis computer system at the Astronomy Data Center, ADC, of the National Astronomical Observatory of Japan (NAOJ).
\end{acknowledgements}

\end{document}